\documentclass[twocolumn]{aastex63}

\usepackage{amsmath,amssymb}

\defcitealias{onodera16}{O16}
\defcitealias{suzuki17}{S17}

\received{}
\revised{}
\accepted{}
\submitjournal{ApJ}

\shorttitle{Dust, gas, and metal in star-forming galaxies at $z\sim3.3$}
\shortauthors{Suzuki et al.}
\graphicspath{{./}{figures/}}
\begin{document}

\title{Dust, gas, and metal content in star-forming galaxies at $z\sim3.3$ revealed with ALMA and Near-IR spectroscopy}

\correspondingauthor{Tomoko L. Suzuki}
\email{suzuki.tomoko@astr.tohoku.ac.jp}

\author[0000-0002-3560-1346]{Tomoko L. Suzuki}
\affiliation{Astronomical Institute, Tohoku University, 6-3, Aramaki, Aoba, Sendai, Miyagi, 980-8578, Japan}
\affiliation{National Astronomical Observatory of Japan, 2-21-1, Osawa, Mitaka, Tokyo, 181-8588, Japan}
\affiliation{Kapteyn Astronomical Institute, University of Groningen, P.O. Box 800, 9700AV Groningen, The Netherlands}

\author[0000-0003-3228-7264]{Masato Onodera}
\affiliation{Subaru Telescope, National Astronomical Observatory of Japan, National Institutes of Natural Sciences (NINS), 650 North A'ohoku Place, Hilo, HI 96720, USA}
\affiliation{Department of Astronomical Science, SOKENDAI (The Graduate University for Advanced Studies), 2-21-1, Osawa, Mitaka, Tokyo, 181-8588, Japan}

\author[0000-0002-2993-1576]{Tadayuki Kodama}
\affiliation{Astronomical Institute, Tohoku University, 6-3, Aramaki, Aoba, Sendai, Miyagi, 980-8578, Japan}

\author[0000-0002-3331-9590]{Emanuele Daddi}
\affiliation{CEA, Laboratoire AIM-CNRS-Universit\'e Paris Diderot, Irfu/SAp, Orme des Merisiers, F-91191 Gif-sur-Y vette, France}

\author[0000-0002-9321-7406]{Masao Hayashi}
\affiliation{National Astronomical Observatory of Japan, 2-21-1, Osawa, Mitaka, Tokyo, 181-8588, Japan}

\author[0000-0002-0479-3699]{Yusei Koyama}
\affiliation{Subaru Telescope, National Astronomical Observatory of Japan, National Institutes of Natural Sciences (NINS), 650 North A'ohoku Place, Hilo, HI 96720, USA}
\affiliation{Department of Astronomical Science, SOKENDAI (The Graduate University for Advanced Studies), 2-21-1, Osawa, Mitaka, Tokyo, 181-8588, Japan}

\author[0000-0003-4442-2750]{Rhythm Shimakawa}
\affiliation{National Astronomical Observatory of Japan, 2-21-1, Osawa, Mitaka, Tokyo, 181-8588, Japan}

\author[0000-0003-3037-257X]{Ian Smail}
\affiliation{Center for Extragalactic Astronomy, Department of Physics, Durham University, South Road, Durham DH1 3LE, UK}

\author[0000-0001-8823-4845]{David Sobral}
\affiliation{Department of Physics, Lancaster University, Lancaster LA1 4YB, UK}
\affiliation{Leiden Observatory, Leiden University, PO Box 9513, 2300 RA Leiden, The Netherlands}

\author[0000-0002-8224-4505]{Sandro Tacchella}
\affiliation{Center for Astrophysics $\vert$ Harvard \& Smithsonian, 60, Garden St, Cambridge, MA 02138, USA}

\author[0000-0002-4937-4738]{Ichi Tanaka}
\affiliation{Subaru Telescope, National Astronomical Observatory of Japan, National Institutes of Natural Sciences (NINS), 650 North A'ohoku Place, Hilo, HI 96720, USA}



\begin{abstract}
We conducted sub-millimeter observations 
with the Atacama Large Millimeter/sub-millimeter Array (ALMA) 
of star-forming galaxies at $z\sim3.3$, 
whose gas-phase metallicities have been previously measured.  
We investigate the dust and gas contents of the galaxies at $z\sim3.3$ 
and study how galaxies are interacting with their circumgalactic/intergalactic medium 
at this epoch by probing their gas mass fractions and gas-phase metallicities. 
Single-band dust continuum emission tracing dust mass and 
the relation between the gas-phase metallicity and gas-to-dust mass ratio 
are used to estimate the gas masses. 
The estimated gas mass fractions and depletion timescales 
are $f_{\rm gas}=$ 0.20--0.75  
and $t_{\rm dep}=$ 0.09--1.55~Gyr, respectively. 
Although the galaxies appear to tightly distribute around the star-forming main sequence at $z\sim3.3$, 
both quantities show a wider spread at a fixed stellar mass than expected from the scaling relation, 
suggesting a large diversity of fundamental gas properties among star-forming galaxies apparently on the main sequence. 
Comparing gas mass fraction and gas-phase metallicity between 
the star-forming galaxies at $z\sim3.3$ and at lower redshifts,
star-forming galaxies at $z\sim3.3$ appear to be more metal-poor 
than local galaxies with similar gas mass fractions.  
Using the gas regulator model to interpret this offset, 
we find that it can be explained by a higher mass-loading factor, 
suggesting that the mass-loading factor in outflows increases at earlier cosmic times. 
\end{abstract}


\section{Introduction} \label{sec:intro}

Molecular gas ($\rm H_2$) %
is one of the fundamental physical quantities 
of galaxies because it is the fuel for star formation.   
It is well known that the gas surface density is correlated 
with the star-formation rate (SFR) surface density 
(the Schmidt--Kennicutt relation; \citealt{schmidt1959,kennicutt98}).
The total gas mass is also connected with the total star-forming activity \citep[e.g.,][]{daddi10,genzel10}.
The typical SFR of star-forming galaxies at a given stellar mass 
appears to monotonically increases with increasing redshifts \citep[e.g.,][]{whitaker12,sobral14,tomczak16}. 
More active star formation in galaxies at higher redshifts 
is considered to be supported by a larger amount of gas 
\citep[e.g.,][]{daddi10,genzel10,geach11,bothwell13_mnras429,tacconi13,birkin20}. 
Investigating the gas contents in galaxies at high redshifts 
is crucial to understand formation and evolution of 
galaxies in the Universe \citep[e.g.,][]{walter16,riechers19}.

Observational studies over the past decade 
revealed the gas properties 
not only for dusty starburst galaxies, such as sub-millimeter bright galaxies (SMGs),   
but also for ultraviolet (UV)/optical--selected star-forming galaxies 
on the stellar mass--SFR relation, 
the so-called ``main sequence'' of star-forming galaxies, 
at $z\gtrsim1$ 
\citep[e.g.,][]{daddi10,genzel10,tacconi10,tacconi13,magdis17}.
The increasing number of galaxies with individual measurements of the gas mass 
in a wide redshift range  
makes it possible to establish the scaling relations for gas mass fraction 
and gas depletion timescale ($\rm =M_{\rm gas}/{\rm SFR}$) 
as a function of redshift, stellar mass, and SFR \citep[e.g.,][]{scoville17,tacconi18,freundlich19,liu19_II}. 
At $z>3$, however, 
the number of UV/optical-selected star-forming galaxies 
with the individual measurements of gas content 
is still small 
(with CO emission lines: \citealt{magdis17,cassata20}, 
and with dust continuum: \citealt{schinnerer16,wiklind19,aravena20}). 
How the gas properties of UV/optical-selected galaxies evolve at $z>3$
are not conclusive yet.

The atomic and/or molecular hydrogen gas mass  
is also said to be correlated with the gas-phase metallicity  
from both observations \citep[e.g.,][]{bothwell13,hunt15,bothwell16_mnras,seko16_alma,brown18}
and cosmological numerical simulations \citep[e.g.,][]{lagos16,torrey19}. 
It has been suggested that the gas mass is more fundamental than the SFR 
to explain the scatter of the mass--metallicity relation 
of star-forming galaxies \citep[e.g.,][]{bothwell13,zahid_apj791,brown18}. 
Indeed, 
more gas-rich star-forming galaxies tend to be more metal-poor 
and more actively forming stars. 
\added{At high redshifts, 
a direct comparison between gas mass 
and gas-phase metallicity is 
limited to a handful of galaxies at $z\sim$\ 1--3 \citep{saintonge13,seko16_alma,shapley20}. 
\citet{seko16_alma} found a trend that the gas mass fraction decreases 
with increasing metallicities at a fixed stellar mass 
for star-forming galaxies at $z\sim1.4$. 
Such a direct comparison between the two quantities  
has not been done at $z>3$.}

Galaxies evolve by interacting with the intergalactic medium (IGM). 
Gas accretes onto galaxies from the outside, 
chemical enrichment proceeds as stars form, 
and gas and metals are ejected from galaxies by outflow \citep[e.g.,][]{bouche10,dave11_1,lilly13,pengmaiolino14,tacchella20}. 
Gas mass fraction and gas-phase metallicity are often used to 
investigate the relative contributions between 
star formation, gas outflow, and inflow \citep{erb08_apj674,cresci10,troncoso14,yabe15_apj,seko16,sanders20}. 
\added{Most of these studies estimated gas mass fractions 
by converting the SFR surface density to gas surface density 
with the Schmidt-Kennicutt relation \citep{erb08_apj674,cresci10,troncoso14,yabe15_apj,sanders20}.}
Given that galaxies are more actively forming stars at higher redshifts, 
they are expected to be more actively interacting with the surrounding IGM 
via outflows and inflows \citep[e.g.,][]{yabe15_apj}. 
At $z>3$, 
it has been suggested that star-forming galaxies are no longer 
in equilibrium \citep[e.g.,][]{mannucci10}, 
where the gas consumption due to star formation and outflows 
is balanced with the gas acquisition by inflows (inflow $=$ star formation $+$ outflow),
due to the intense gas inflows onto galaxies in the early Universe. 
Estimating both the gas mass and gas-phase metallicity 
for star-forming galaxies at $z>3$ 
allows tests of whether 
galaxies at $z>3$ are out of equilibrium or not. 
Several methods are commonly used to estimate gas masses. 
The first one is using CO emission line fluxes \citep[e.g.,][]{daddi10,genzel10,tacconi10,tacconi13}. 
This method has uncertainties on 
the CO-to-$\rm H_{2}$ conversion factor, which changes depending on metallicity \citep{genzel12}
and on the CO excitation states when using higher-$J$ CO lines \citep[e.g.,][]{daddi15,riechers20}.  
Furthermore, observations of CO lines for galaxies at high redshifts 
are observationally expensive. 
The second one is converting 
a dust mass to a gas mass with an assumed gas-to-dust mass ratio \citep[e.g.,][]{santini14,bethermin15}. 
Because the gas-to-dust mass ratio depends on the metallicity \citep{leroy11,remy-ruyer14}, 
metallicity measurements are crucial to estimate the gas mass accurately. 
Gas masses can also be estimated with an empirically calibrated relation between 
a single-band dust continuum flux at the Rayleigh-Jeans (R-J) tail 
and gas mass \citep[e.g.,][]{scoville14,scoville16,groves15}. 
These scaling relations are calibrated with local galaxies 
or with local galaxies and SMGs up to $z\sim2$. 
In this method, the gas-to-dust mass ratio is included in the conversion factor, 
and thus, is not needed to be considered. 
It remains unclear whether the empirical calibration methods are applicable 
to galaxies at $z>3$ 
or how much scatter there is in the relationships. 
Given that dust continuum observations take much less time as compared to the CO observations, 
using dust continuum as a tracer of gas has an advantage to increase the number of galaxies 
at higher redshifts with individual gas estimates, 
but these will only be reliable when precise metallicities are also available. 
\added{Metallicity measurements based on rest-frame optical emission lines 
for dustier star-forming galaxies 
are thought to have larger uncertainties 
due to strong dust obscuration \citep[e.g.,][]{santini10}. 
\citet{herrera-camus18} reported a discrepancy between metallicities 
derived with rest-frame optical emission lines and far-infrared (FIR) 
fine-structure lines for local (Ultra) Luminous Infrared Galaxies ((U)LIRGs).}

In this paper, 
we present the results from sub-millimeter observations 
with the Atacama Large Millimeter/sub-millimeter Array (ALMA) 
of star-forming galaxies at $z=$\ 3--4. 
High quality near-infrared (NIR) spectra obtained with Keck/MOSFIRE \citep{mclean10,mclean12}
are available for all of the targets 
and their gas-phase metallicities are already measured \citep{onodera16,suzuki17}. 
By observing the dust continuum emission, 
we can estimate their dust masses and convert them to gas masses 
using the relation between the metallicity and gas-to-dust mass ratio. 
We then investigate the gas properties, 
namely, gas mass fractions and gas depletion timescales, 
of star-forming galaxies at $z=$\ 3--4. 
Comparing the gas contents with the gas-phase metallicities, 
we aim to understand how star-forming galaxies at this epoch interact 
with their surrounding IGM via gas inflows and outflows.

This paper is organized as follows. 
In Section~\ref{sec:obs}, we introduce our parent sample 
of star-forming galaxies at $z=$\ 3--4 and describe the observations conducted with ALMA. 
We also describe the reduction and analysis for the obtained data 
and stacking analysis. 
In Section~\ref{sec:physicalquantity}, 
we present our estimates of the physical quantities, such as 
gas-phase metallicity, ionization parameter, and gas mass. 
In Section~\ref{sec:results}, 
we show our results on the dust and gas properties 
of the star-forming galaxies at $z=$\ 3--4 and discuss their metallicity dependencies.  
We also compare our observational results with a gas regulator model 
to discuss how star-forming galaxies at this epoch interact with their surrounding IGM. 
We summarize this paper in Section~\ref{sec:summary}.

Throughout of this paper, 
we assume the cosmological parameters of $\rm \Omega_m = 0.3$, $\rm \Omega_{\Lambda} = 0.7$, and $H_{\rm 0} = 70\ {\rm km\ s^{-1}\ Mpc^{-1}}$. 
We use a Chabrier initial mass function (IMF; \citealt{chabrier03}).

\section{Observation and reduction} \label{sec:obs}

\subsection{Spectroscopically confirmed galaxies at $z=$3--4}
Our parent sample is constructed 
from the two different studies 
of star-forming galaxies at $3 < z < 4$ in the COSMOS field 
using NIR spectroscopy with Keck/MOSFIRE. 
One study is based on 
a spectroscopic and photometric redshift selection 
\citep[Section~\ref{subsec:o16}]{onodera16}, 
while the other study is based on narrow-band selection 
\citep[Section~\ref{subsec:s17}]{suzuki17}. 
The parent sample from both studies consists of 53 galaxies with $z_{\rm spec}\sim$\ 3.0--3.8.

\subsubsection{UV-selected galaxies}\label{subsec:o16}
In \citet[][hereafter O16]{onodera16},  
targets for spectroscopic observation were originally selected from 
the zCOSMOS-Deep redshift catalog \citep{lilly07}
and the 30-band COSMOS photometric redshift catalog \citep{mccracken12,ilbert13}. 
\citetalias{onodera16} conducted {\it H}- and {\it K}-band spectroscopy and 
confirmed 43 galaxies at $z_{\rm spec}=$\ 3.0--3.8 based on the rest-frame optical emission lines. 
The confirmed star-forming galaxies span a stellar mass range of 
$\rm log(M_*/M_\odot) \sim$\ 8.5--11.0 and distribute around the star-forming main sequence 
at $z\sim$ 3.3 \citepalias{onodera16}.

\subsubsection{[OIII] emission line galaxies}\label{subsec:s17}
In \citet[][hereafter S17]{suzuki17}, 
targets for spectroscopic observation were selected from 
a catalog of narrow-band(NB)-selected [{\sc Oiii}]$\lambda$5007 emission line galaxies at $z=3.23$, 
obtained by the High-Z Emission Line Survey 
(HiZELS; \citealt{best13, sobral13, khostovan15}). 
\citetalias{suzuki17} conducted {\it H}- and {\it K}-band spectroscopy 
and confirmed ten [{\sc Oiii}] emitters at $z_{\rm spec}=$\ 3.23--3.28. 
\added{The stellar mass range of the confirmed [{\sc Oiii}] emitters 
is $\rm log(M_*/M_\odot) \sim$\ 9.1--10.2.}
The [{\sc Oiii}] emitters follow the star-forming main sequence at $z\sim3.2$  
and the mass--metallicity relation established by \citetalias{onodera16} \citepalias{suzuki17}.

\subsection{ALMA Band-6 observation}\label{subsec:almaobs}
We selected galaxies with 
$\rm log(M_*/M_\odot) \ge 10$ and $\ge 3\sigma$ detection of 
[{\sc Oiii}]$\lambda$5007, H$\beta$, or [{\sc Oii}]$\lambda$3727 emission lines 
from the parent sample 
as targets for the ALMA observations. 
We excluded two galaxies classified as the active galactic nuclei (AGNs) in \citetalias{onodera16}.
One has an X-ray counterpart detected with {\it Chandra}. 
The other shows strong [Ne{\sc iii}]$\lambda$3869 emission and a high [{\sc Oiii}]$\lambda$4363/H$\gamma$ ratio, 
which is likely to be powered by the AGN \citepalias{onodera16}.
As a result, 
the sample for the ALMA observations consists of 
12 galaxies, two of which are [{\sc Oiii}] emitters 
from \citetalias{suzuki17} (Table~\ref{table:obssummary}).

Although the potential AGNs were excluded from our ALMA targets, 
we found that one of the ALMA targets, 192129, 
is detected in X-ray with {\it Chandra} \citep{elvis09,civano12,civano16} 
and included in the X-ray-selected AGN catalog of \citet{kalfountzou14} as a type-2 AGN. 
The optical--NIR spectral energy distribution (SED) of this source is not peculiar as compared to 
the other galaxies \citepalias[][and as shown in the best-fit SEDs in Appendix~\ref{sec:appendix1}]{onodera16} 
and its H$\beta$ emission line is narrow, 
which is likely to be consistent with the classification by \citet{kalfountzou14}. 
Although we expect that the optical--NIR emission is dominated by 
emission from the host galaxy, 
the physical quantities, 
such as stellar mass, SFR, gas-phase metallicity, 
and ionization parameter (Section~\ref{subsec:magphys} and \ref{subsec:ism}), 
may be affected by the emission from the AGN. 
On the other hand, 
the dust continuum observed at ALMA Band-6 ($\lambda_{\rm rest}\sim280\ \mu {\rm m}$)  
is expected to be dominated by cold dust emission from star-forming regions ($T\sim$ 20--40~K).
We do not exclude this source in the following analyses 
but distinguish it from the other sources on each figure.

Our ALMA Cycle~6 program with Band-6 was  
conducted during December 2018 -- March 2019 
(2018.1.00681.S, PI: T. Suzuki). 
Frequencies of four spectral windows are slightly different among the targets   
depending on their spectroscopic redshifts (between $221.9$ GHz and $254.4$ GHz). 
We set the frequencies of the spectral windows so that 
we can cover the CO(9--8) line ($\nu_{\rm rest} = 1036.9$ GHz) with one of the four spectral windows.  
The effective bandwidth of each spectral window is $~1.875$ GHz. 
The data were taken with the Time Domain Mode (TDM). 
The total on-source time is $\sim$5--90~min 
depending on the stellar mass, SFR, and gas-phase metallicity of the targets as summarized in Table~\ref{table:obssummary}.  

The brightest source at 1.3~mm, 208681, 
appears to be detected with the CO(9--8) line. 
\added{Given that quasar host galaxies tend to 
have more extreme CO excitation states than 
normal star-forming galaxies \citep{carilliwalter13}, 
the CO(9--8) line detection 
may suggest that this source hosts an AGN.}
We will discuss the CO(9--8) line of this source in a forthcoming paper 
(Suzuki et al. 2020 in preparation). 
\added{We note that} 
the contribution of the CO(9--8) line to the dust continuum flux is negligible.

We used the Common Astronomy Software Application package ({\sc casa}; \citealt{CASA})
to calibrate the raw data. 
We run the {\sc clean} algorithm with natural weighting.
When there are sources detected with $\ge 5\sigma$ level, 
we run {\sc clean} again by masking the sources.  
\added{Because the synthesized beam sizes are slightly different among the targets, 
we created the uv-tapered maps for 
some of the sources to conduct flux measurement 
under similar beam sizes. 
The average beam size of 12 ALMA maps is $1.''52 \times 1.''32$.} 

\added{
We used {\sc imfit} to fit a 2D Gaussian to each target. 
The central position is fixed at the centroid determined in the {\it Ks}-band image 
from UltraVISTA\footnote{\url{https://irsa.ipac.caltech.edu/data/COSMOS/index_cutouts.html}}. 
Our detection criterion is that the peak flux obtained by {\sc imfit} has 
$>3\sigma$ significance. 
As a result, six out of 12 galaxies satisfy this criterion as summarized in Table~\ref{table:obssummary}.
We ran {\sc imfit} several times with different parameter settings to check the robustness 
of the obtained fluxes. 
We confirmed that 
the fitting results for the six detected sources are not affected by the parameter settings. 
In the following sections, we use peak fluxes measured with {\sc imfit} 
as the total fluxes of the detected sources.  
The obtained peak fluxes broadly agree with 
the aperture fluxes ($r=1.5$~arcsec) measured at the position of 
the {\it Ks}-band centroid within the uncertainties, 
which suggests that our targets are not spatially resolved 
in the ALMA maps. 
As for the non-detected sources, 
we assigned $3\sigma$ upper limit fluxes. 
The measured fluxes and limits are summarized in Table~\ref{table:obssummary}.
The listed fluxes are corrected for the primary beam.
Because the ALMA targets are located at the center of each ALMA map,  
the primary beam correction is less than 1~\% for all of the targets.}

\begin{figure*}[t]
\centering\includegraphics[width=1.0\textwidth]{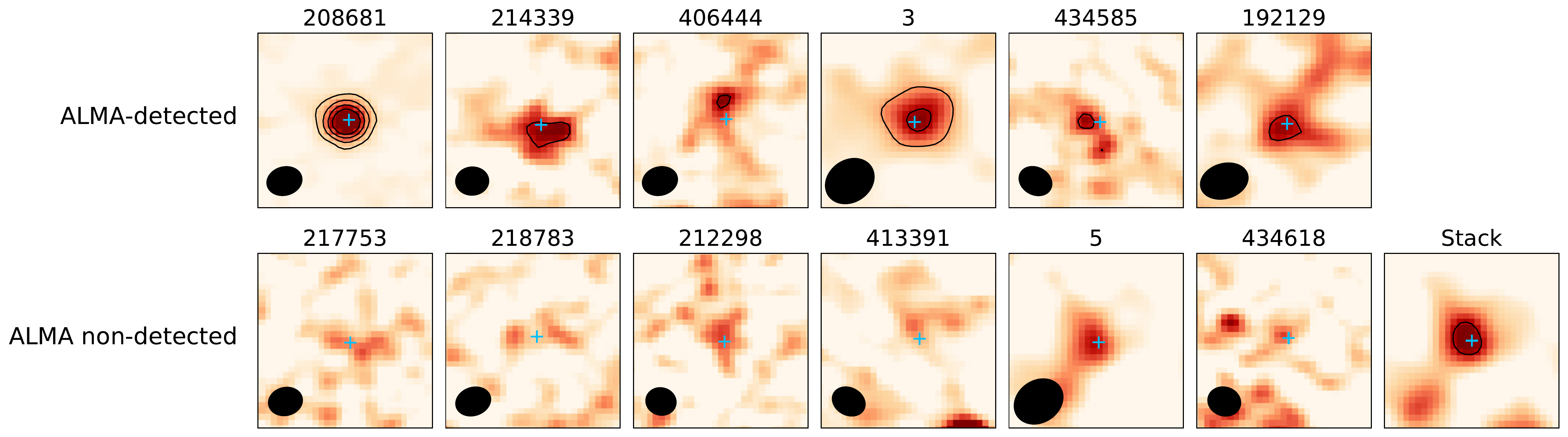}
\caption{
ALMA Band-6 images before tapering of 12 targets (image size: $5''\times5''$).
A black circle shows the beam size of each image. 
Black contours correspond to 4$\sigma$, 8$\sigma$, $12\sigma$, and 16$\sigma$. 
A plus mark represents the centroid determined in the {\it Ks}-band image. 
%Two out of six
\added{Some of the}
ALMA detected sources, \added{such as} 406444 and 434585, 
show a larger spatial offset between the dust continuum emission and 
the {\it Ks}-band centroid than the other detected sources. 
This \added{is probably} due to the lower signal-to-noise ratios of their dust continuum emission. 
The stacked image of the five individually ALMA non-detected sources with 
$\rm log(M_*/M_\odot)=$ 10.0--10.4 is also shown.  
The central position is shown with a plus mark. 
The stacked emission is detected at $5\sigma$. 
}
\label{fig:image}
\end{figure*}

Figure~\ref{fig:image} shows the ALMA maps of our target galaxies 
together with the {\it Ks}-band centroids. 
\added{Some of the ALMA-detected sources, such as 406444 and 434585, show a clear spatial offset 
(up to $\sim0.5$~arcsec)
between the dust continuum emission and the {\it Ks}-band centroid. 
Their relatively large positional offsets are probably due to 
the lower signal-to-noise ratios of their dust continuum emission. 
Indeed, among the ALMA-detected sources, 
we found a trend that the positional offset between the dust emission peak 
and the {\it Ks}-band centroid becomes larger with 
decreasing signal-to-noise ratio of the dust continuum emission.}

\begin{table*}[]
\caption{Summary of the targets of this study and ALMA Band-6 observations.}
    \begin{center}
    \begin{tabular}{ccccccccc}\hline
    ID$^{\rm a}$ & R.A.$^{\rm a}$ & DEC.$^{\rm a}$ & $z_{\rm spec}$ & Exp. time$^{\rm b}$ & Central freq.$^{\rm c}$ &  RMS level$^{\rm d}$ & $S_{\rm 1.3mm}$ $^{\rm e}$ & Reference \\  
     & [deg] & [deg] & & [min] & [GHz] &  [mJy $\rm beam^{-1}$] & [mJy] &  \\ \hline
     208681 & 149.90551 & 2.353990 & 3.267 & 5 & 232.1 & 0.05 & $1.02\pm0.05$  &  \citetalias{onodera16} \\  
     214339 & 150.31607 & 2.372240 & 3.609 & 5  & 233.0 & 0.05 & $0.30 \pm 0.05$ & $''$ \\
     406444 & 150.33032 & 2.072270 & 3.304 & 5 & 232.1 & 0.04 & $0.12\pm0.04$ & $''$ \\
          3 & 149.97513 & 1.69375 & 3.230 & 41 & 235.6 & 0.01 & $0.11\pm0.01$ & \citetalias{suzuki17} \\
     434585 & 149.84702 & 2.373020 & 3.363 & 11 & 230.9 & 0.03 & $0.11 \pm 0.03$  & \citetalias{onodera16} \\
     $192129^{\rm f}$ & 150.30078 & 2.300540 & 3.495 & 49 & 237.3 & 0.01 & $0.05 \pm 0.01$ &  $''$ \\
     217753 & 149.89451 & 2.383700 & 3.254 & 5 & 232.1 & 0.05 & $< 0.15$ &  $''$ \\     
     218783 & 149.92082 & 2.387060  & 3.297 & 5 & 232.1 & 0.04 & $< 0.11$ &  $''$ \\
     212298 & 150.34268 & 2.365390 & 3.108 & 5  & 246.2 & 0.04 & $< 0.11$ & $''$ \\
     413391 & 149.78424 & 2.452890 & 3.365 & 11 & 230.9 & 0.03 & $< 0.10$ & $''$ \\
     5 & 149.95568 & 1.68044 & 3.241 & 53 & 235.6 & 0.01 & $< 0.04$ &  \citetalias{suzuki17} \\
     434618 & 149.89213 & 2.414710 & 3.285 & 88 & 233.5 & 0.01 & $< 0.03$ & \citetalias{onodera16} \\  \hline
    \end{tabular}
\end{center}
\tablecomments{
    $^{\rm a}$Object IDs and coordinates are extracted from the original papers. 
    $^{\rm b}$\added{On-source observing time} for Band-6 observations. 
    $^{\rm c}$Central frequency of the four spectral windows. 
    $^{\rm d}$Measured in the tapered maps.
    $^{\rm e}$Measured in the tapered maps with {\sc imfit} \added{and corrected for the primary beam}. 
    3$\sigma$ upper limits are shown for the ALMA non-detected sources. 
    $^{\rm f}$X-ray detected source (Section~\ref{subsec:almaobs}). 
    }
    \label{table:obssummary}
\end{table*}

\subsection{Stacking analysis}\label{subsec:stack}

We stacked the Band-6 images of the ALMA non-detected sources 
to investigate their average dust continuum flux. 
As a result of SED fitting in Section~\ref{subsec:magphys}, 
%including ALMA Band-6 data 
one source, 434618, turns out to be only $\rm log(M_*/M_\odot)=9.39_{-0.01}^{+0.12}$. 
\added{This is probably due to using a different SED fitting code 
and/or different photometric catalog with deeper NIR and {\it Spitzer} data 
from the previous estimate \citepalias{onodera16}.}
Because the stellar mass of 434618 is $\gtrsim 0.6$~dex smaller than those of the other non-detected sources, 
we excluded this source 
so that we can investigate the average dust and gas properties of star-forming galaxies with similar stellar masses of 
$\rm log(M_{*}/M_\odot)=$\ 10.0--10.4.

We cut out the tapered $20''\times20''$ ALMA images centered on 
the {\it Ks}-band position. 
Then, we stacked the cutout images by weighting with the RMS levels in the tapered maps (Table~\ref{table:obssummary}).  
The stacked image is shown in Figure~\ref{fig:image}. 
\added{We measured the total flux of the stacked image with {\sc imfit} as done in 
Section~\ref{subsec:almaobs}. 
The obtained flux is $0.06\pm0.01\ {\rm mJy}$, which satisfies our detection 
criterion.}

\subsection{SED fitting}\label{subsec:magphys}

We conducted SED fitting  
including the dust continuum flux or limit at $1.3$~mm obtained with ALMA. 
We used the SED fitting code {\sc magphys} \citep{dacunha08,dacunha15,battisti19}, 
which can fit the SEDs from the optical to radio 
wavelengths consistently. 
{\sc magphys} combines the emission by stellar populations 
with the attenuation and emission by dust 
in galaxies based on the energy balance technique. 
We used the updated version of {\sc magphys} for galaxies at high redshifts \citep{dacunha15}.

{\sc magphys} adopts stellar population synthesis models of \citet{bc03} 
with the \citet{chabrier03} IMF. 
The metallicity range is set to be from 0.2 to 2 times solar. 
Star formation history is parameterized as a continuous delayed exponential function 
i.e., the star formation rate rises at the earlier epoch and 
then declines exponentially with a certain timescale between 0.075 and 1.5 $\rm Gyr^{-1}$.   
The age is randomly drawn between 0.1 and 10 Gyrs. 
{\sc magphys} also includes star bursts of random duration 
and amplitude to account for the stochasticity on star formation history.  
The current SFR is determined by averaging SFH over last 
\added{100~Myr}. 
As for the dust attenuation, 
{\sc magphys} uses the two-component model of \citet{charlotfall00}. 
A number of tests of the application of {\sc magphys} 
to dust-obscured galaxies, 
including simulated galaxies from EAGLE, 
are discussed in \citet{dudzeviciute20}. 

{\sc magphys} takes into account four main dust components, 
namely, a polycyclic aromatic hydrocarbons, hot dust at mid-infrared, 
warm dust, and cold dust in the thermal equilibrium. 
The warm and cold dust components in the thermal equilibrium emit 
as modified black bodies with emissivity index $\beta$ of 1.5 for the warm components 
and 2 for the cold components \citep{dacunha15}. 
{\sc magphys} assumes a dust mass absorption coefficient at $850\mu$m of $\kappa_{\rm abs}=0.77\ {\rm g^{-1}\ cm^2}$.

We combined the flux densities at $1.3$~mm from ALMA 
with the optical-NIR broad-band photometries ($u,B, V, r, i_p, z_{pp}, Y, J, H, Ks$, 3.6, 4.5, 5.8, and 8.0 $\mu$m)  
from the COSMOS2015 catalog 
\citep{cosmos2015}. 
Because {\sc magphys} does not include emission lines from the ionized gas, 
we subtracted the emission line fluxes measured with the NIR spectra 
from the {\it H}-band ([{\sc Oii}]) and {\it Ks}-band ([{\sc Oiii}] doublet, H$\beta$) fluxes. 
We took into account the upper limits for the optical-NIR photometries 
by giving 0 to the flux column and 
a $3\sigma$ value to the photometric error column according to \citet{dacunha15}. 
As for the 1.3~mm flux of the ALMA non-detected sources, 
we gave a $1.5\sigma \pm 1\sigma$ value   
as done in \citet{dudzeviciute20}.  
Using a $1.5\sigma \pm 1\sigma$ value 
provides a better weighting of the sub-millimeter constraint 
in the best-fit model returned by {\sc magphys} 
than using a $3\sigma$ upper limit. 
\added{The derived physical parameters, such as stellar masses 
and SFRs, do not significantly change 
depending on the adopted flux and error values \citep{dudzeviciute20}.}

We also conducted SED fitting for the stacked sample with the obtained $1.3$~mm flux 
in Section~\ref{subsec:stack}. 
When taking an average of the optical--NIR photometries, 
we used the same weights as used in the ALMA image stacking (Section~\ref{subsec:stack}).

The best-fit SEDs of the individual galaxies and the stacked sample are shown in Appendix~\ref{sec:appendix1}. 
We use the median values of the probability distribution function (PDF) 
for stellar mass, SFR, and dust mass in the following analyses. 
These physical quantities are summarized in Table~\ref{tab:basicquantity}. 
The uncertainties correspond to the 16--84th percentile values of the PDF. 
As for the dust masses of the ALMA non-detected sources, 
we use the 97.5th percentile values of the PDF as the upper limits.

In order to evaluate whether the upper limits on the dust masses 
are reasonable, 
we estimated dust mass upper limits with a different method. 
We calculated a ratio between the dust mass and the luminosity density 
at $997.6$~GHz in the rest-frame, 
${\rm M_{dust}}/L_{\rm 997.6~GHz}$, for each detected source.
We then converted the $3\sigma$ upper limits of 1.3~mm fluxes to the dust mass upper limits 
with the median ${\rm M_{dust}}/L_{\rm 997.6~GHz}$ ratio. 
The difference of the rest-frame frequencies among the sources is corrected for assuming $L_{\nu} \propto \lambda^{-3.7}$. 
The estimated dust mass upper limits are similar as 
the 97.5th percentile values of the PDF from {\sc magphys}.

%% systematic uncertainty?
One of the uncertainties on the dust mass is the assumed dust mass absorption coefficient, 
$\kappa_{\rm abs}$.
It is reported that dust masses obtained with {\sc magphys} 
are lower by a factor of two as compared to 
those estimated based on the \citet{draineli07} models, 
which assume the smaller dust mass absorption coefficient of $\kappa_{\rm abs}=0.38\ {\rm g^{-1}\ cm^2}$ \citep{hunt19}.

\begin{table*}[]
\caption{Summary of the physical quantities of the star-forming galaxies at $z\sim3.3$ and the stacked sample.}
    \begin{center}
    \begin{tabular}{cccccccc}\hline
      ID  & $\rm log(M_*/M_\odot)^{a}$ & $\rm log(SFR)^{a}$ & $\rm 12+log(O/H)^{\rm b}$ & log($q$) & $\delta_{\rm GDR}$ & $\rm log(M_{\rm dust}/M_\odot)^{a,c}$ & $\rm log(M_{gas}/M_\odot)$  \\  
          &  & [$\rm M_\odot\ yr^{-1}$] & & [$\rm cm\ s^{-1}$] & & & \\ \hline
      208681  & $10.82_{-0.03}^{+0.00}$ & $2.10_{-0.01}^{+0.08}$ & $8.59_{-0.05}^{+0.04}$ & $7.64\pm0.03$ & $126_{-14}^{+13}$ & $8.89_{-0.08}^{+0.11}$ & $11.29_{-0.09}^{+0.11}$ \\
      214339  & $10.48_{-0.05}^{+0.04}$ & $1.76\pm0.13$ & $8.30_{-0.12}^{+0.09}$ & $7.82_{-0.06}^{+0.05}$ & $264_{-70}^{+54}$ & $8.22_{-0.16}^{+0.17}$ & $10.94_{-0.20}^{+0.19}$ \\
      406444  & $10.87_{-0.01}^{+0.08}$ & $2.31_{-0.11}^{+0.08}$ & $8.39_{-0.09}^{+0.08}$ & $7.77_{-0.09}^{+0.08}$ & $211_{-49}^{+40}$ & $7.64_{-0.13}^{+0.12}$ & $10.26_{-0.17}^{+0.15}$  \\
      3       & $10.43_{-0.08}^{+0.06}$ & $1.73_{-0.15}^{+0.18}$ & $8.40_{-0.06}^{+0.05}$ & $7.69\pm0.06$ & $205_{-29}^{+26}$ & $7.71_{-0.13}^{+0.15}$ & $10.32_{-0.14}^{+0.16}$ \\
      434585  & $10.13_{-0.12}^{+0.04}$ & $1.73_{-0.01}^{+0.17}$ & $8.47_{-0.14}^{+0.10}$ & $7.69\pm0.25$ & $172_{-64}^{+46}$ & $7.67_{-0.22}^{+0.20}$ & $10.21_{-0.28}^{+0.23}$ \\
      192129  & $10.45_{-0.00}^{+0.12}$ & $1.35_{-0.08}^{+0.00}$ & $8.41_{-0.08}^{+0.07}$ & $7.83_{-0.04}^{+0.03}$ & $199_{-41}^{+33}$ & $7.40_{-0.28}^{+0.18}$ & $10.00_{-0.29}^{+0.19}$ \\
      
      217753  & $10.39_{-0.06}^{+0.05}$ & $1.67_{-0.18}^{+0.12}$ & $8.57_{-0.05}^{+0.04}$ & $7.55\pm0.06$ & $131_{-17}^{+15}$ & $< 8.00$ & $< 10.42$ \\
      218783  & $10.12_{-0.07}^{+0.08}$ & $1.70_{-0.17}^{+0.16}$ & $8.42_{-0.07}^{+0.06}$ & $7.67\pm0.03$ & $197_{-35}^{+30}$ & $< 7.84$ & $ < 10.43$ \\
      212298  & $10.38_{-0.08}^{+0.14}$ & $1.74_{-0.20}^{+0.26}$ & $8.39_{-0.08}^{+0.07}$ & $7.76_{-0.04}^{+0.03}$ & $213_{-43}^{+35}$ & $< 7.84$ & $< 10.47$ \\
      413391  & $10.08_{-0.01}^{+0.00}$ & $1.88\pm0.00$ & $8.33_{-0.10}^{+0.08}$ & $7.69_{-0.07}^{+0.06}$ & $246_{-56}^{+47}$ & $< 7.73$ & $< 10.42$ \\
      5       & $10.14_{-0.05}^{+0.09}$ & $1.37\pm0.15$ & $8.37\pm0.05$ & $7.59\pm0.07$ & $220_{-27}^{+25}$ & $< 7.33$ & $< 9.97$ \\
      434618  & $9.39_{-0.01}^{+0.12}$  & $1.18_{-0.09}^{+0.00}$ & $8.26_{-0.09}^{+0.08}$ & $7.78_{-0.04}^{+0.03}$ & $284_{-58}^{+49}$ & $< 7.28$ & $< 10.04$ \\ \hline
      stack$^{\rm d}$ &  $10.18_{-0.06}^{+0.05}$ & $1.52_{-0.14}^{+0.15}$ & $8.38\pm0.05$ & $7.62\pm0.06$ & $216\pm27$  & $7.33_{-0.15}^{+0.17}$  & $9.97_{-0.16}^{+0.18}$ \\ \hline         
    \end{tabular}
    \end{center}
     \tablecomments{
    $^{\rm a}$Median value of the PDF obtained from {\sc magphys}. 
    Error bars correspond to the 16th--68th percentiles.
    $^{\rm b}$Estimated using an empirical calibration method by \citet{curti16}. 
    $^{\rm c}$The 97.5th percentile values from {\sc magphys} are given as the upper limits. 
    $^{\rm d}$Stacking result for the five ALMA non-detected sources with $\rm log(M_*/M_\odot)=$\ 10.0--10.4 (Section~\ref{subsec:stack}).
    }
    \label{tab:basicquantity}
\end{table*}

\section{Analysis} \label{sec:physicalquantity}

\subsection{Gas-phase metallicity and ionization parameter}\label{subsec:ism}

We recalculated the gas-phase metallicities 
\added{with the following four relations, which are locally calibrated 
in \citet{curti16}: 

\begin{eqnarray}
    &{\rm log}\ {R_{\rm 2}}& = 0.418 - 0.961x - 3.505x^2 - 1.949x^3, \\ 
    &{\rm log}\ {R_{\rm 3}}& = -0.277 - 3.549x - 3.593 x^2 - 0.981x^3, \\
     &{\rm log}\ {O_{\rm 32}}& = -0.691 - 2.944x -1.308x^2, \\
     &{\rm log}\ {R_{\rm 23}}& = 0.527 - 1.569x - 1.652x^2 - 0.421x^3, 
\end{eqnarray}
    
\noindent 
where $R_{\rm 2}=$ [{\sc Oii}]/H$\beta$, $R_{\rm 3}=$ [{\sc Oiii}]$\lambda$5007/H$\beta$, 
$O_{\rm 32}=$ [{\sc Oiii}]$\lambda$5007/[{\sc Oii]}, 
$R_{\rm 23}=$ ([{\sc Oiii}]$\lambda\lambda$4959,5007 + [{\sc Oii}])/H$\beta$, 
and $x$ is $\rm 12+log(O/H)$ normalized to the solar value.}
The emission line fluxes of the sources are available in \citetalias{onodera16} and \citetalias{suzuki17}. 

We also estimated the ionization parameter, ${\rm log}(q)$, for the galaxies observed with ALMA 
as done in \citetalias{onodera16}. 
The ionization parameter is described as the ratio of the number of the ionizing photons 
and the hydrogen atoms to be ionized. 
The definition of $q$ is as follows:

\begin{equation}
q = \frac{Q_{\rm H^0}}{4\pi R_s^2n_{\rm n_H}}, 
\end{equation}

\noindent
where $Q_{\rm H^0}$ is the flux of the ionizing photons produced by the existing stars above the Lyman limit, 
$R_s$ is the Str\"omgren radius, and $n_{\rm H}$ is the local density 
of hydrogen atoms \citep{kewleydopita02}.

We use the following relation by \citet{KK04} to estimate the ionization parameter 
from the [{\sc Oiii}]$\lambda\lambda$4959,5007/[{\sc Oii}] ratio ($O_{\rm 32}$) and gas-phase metallicity;

\begin{eqnarray}
    {\rm log}(q) = &\{&32.81 - 1.153y^2 + [{\rm 12+log(O/H)}] \nonumber \\ \nonumber
    &\times& (-3.396-0.025y+0.1444y^2)\} \\ \nonumber
    &\times& \{4.63-0.3119y-0.163y^2 + [{\rm 12+log(O/H)}] \\ 
    &\times& (-0.48 + 0.0271y+0.02037y^2)\}^{-1}, 
\end{eqnarray}

\noindent 
where $y={\rm log}\ O_{\rm 32}$.

\subsection{$M_*$ -- SFR and $M_*$ -- gas-phase metallicity diagram}

\begin{figure*}[t]
\begin{minipage}[cbt]{0.49\textwidth}
    \centering
    \includegraphics[width=0.9\textwidth]{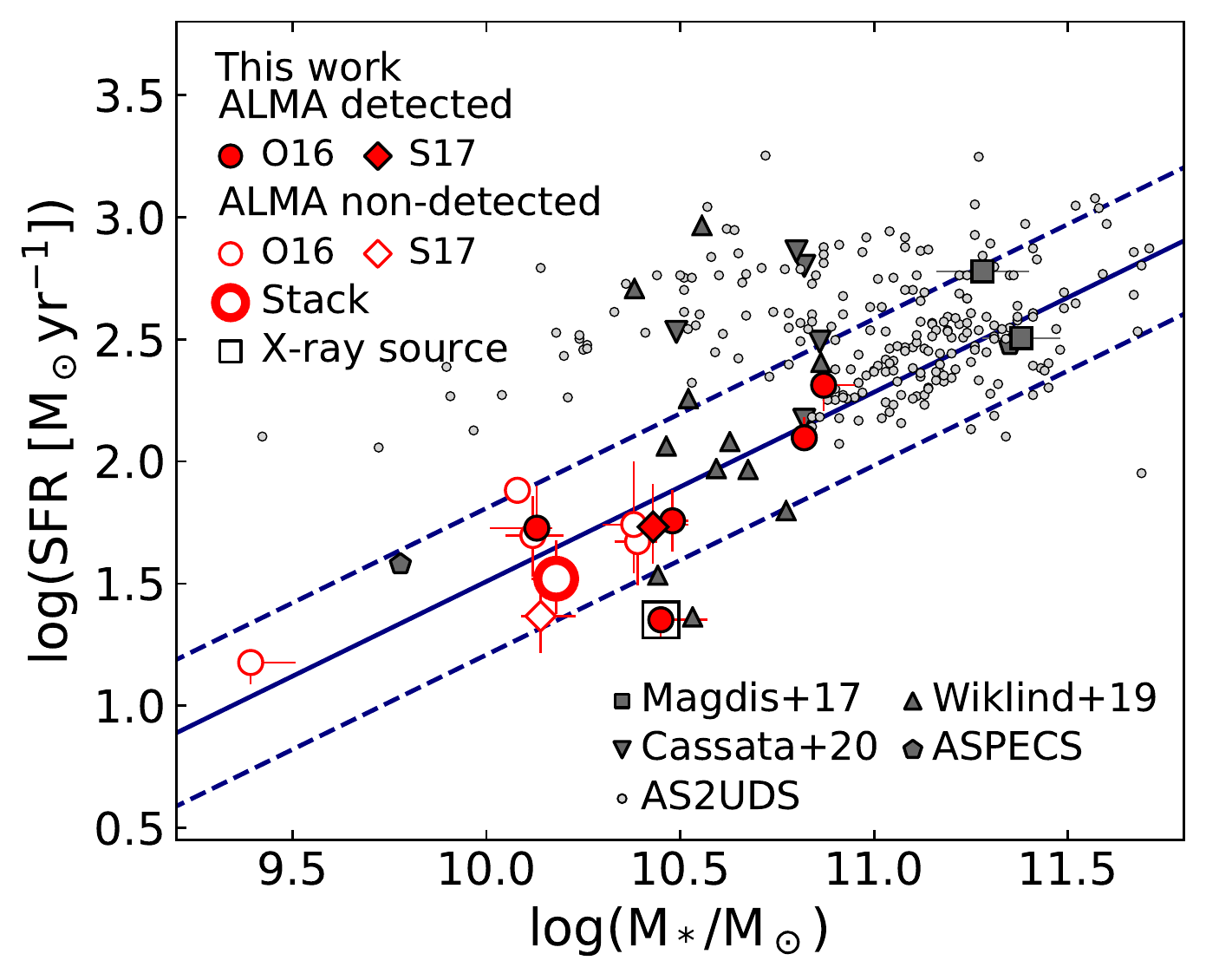}
\end{minipage}
\begin{minipage}[cbt]{0.49\textwidth}
    \centering
    \includegraphics[width=0.9\textwidth]{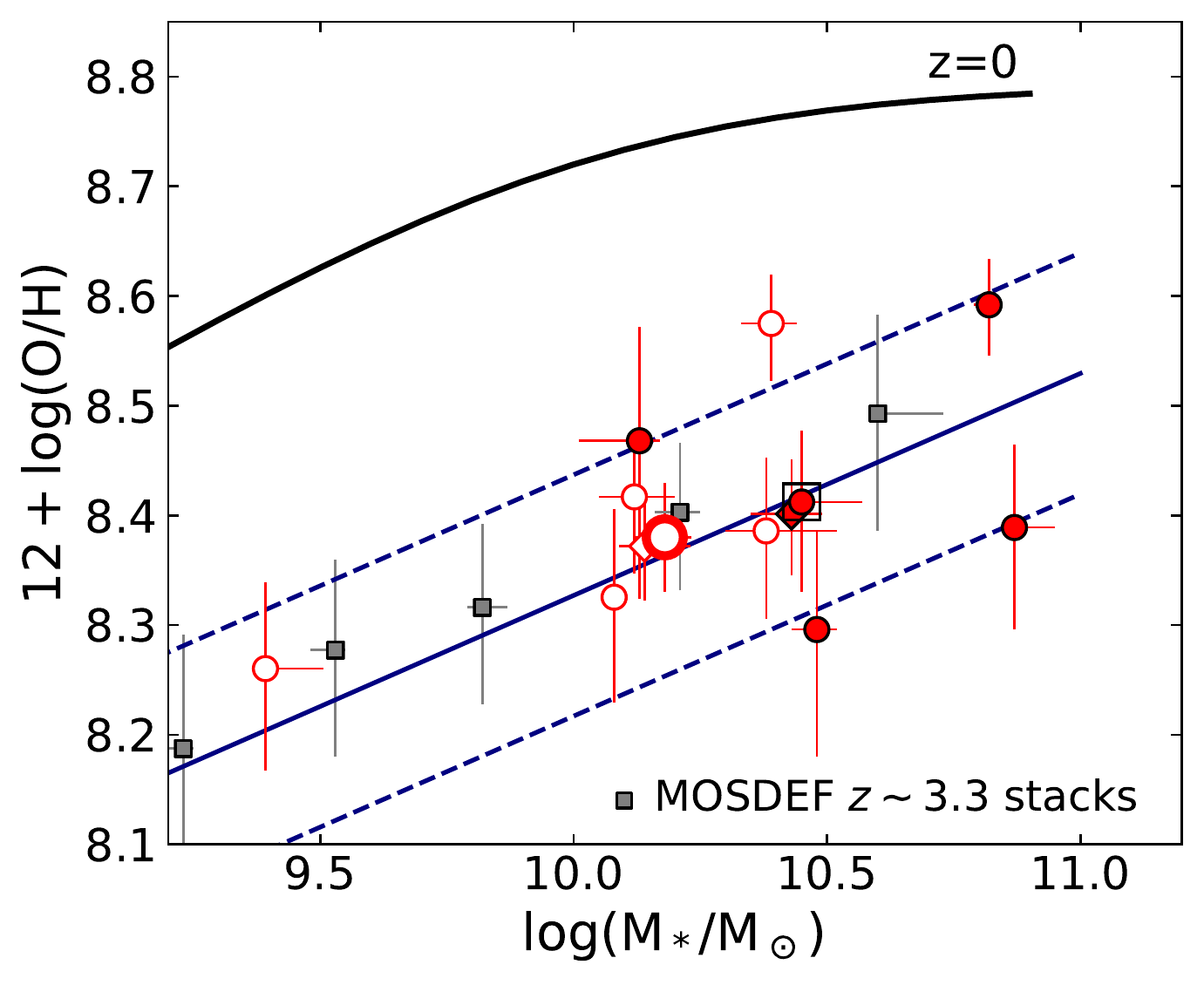}
\end{minipage}
\caption{(Left) Stellar mass--SFR relation for the star-forming galaxies at $z=$\ 3--4 observed with ALMA 
together with star-forming galaxies and SMGs at $z=$\ 3--4 from the literature (Section~\ref{subsec:comparisonsample}).   
The solid line shows the star-forming main sequence at $z=3.3$ from \citet{speagle14}. 
The dashed lines represent $\pm0.3\ {\rm dex}$ from the main sequence.
(Right) Stellar mass versus gas-phase metallicity diagram.  
\added{We show the stacking results from the MOSDEF survey \citep{sanders20} for comparison.}
The thick solid line shows the mass--metallicity relation at $z=0$ from \citet{curti20}. 
The black solid line shows the best-fitted relation for our parent sample at $z\sim3.3$, 
and the dashed lines represent its scatter of $0.11$~dex. 
The galaxies observed with ALMA distribute around the star-forming main sequence 
and the mass--metallicity relation.
They are not biased in terms of 
the star-forming activity and gas-phase metallicity. 
}
\label{fig:msmz}
\end{figure*}

Figure~\ref{fig:msmz} shows the star-forming main sequence and the mass--metallicity relation 
for the star-forming galaxies at $z\sim3.3$. 
In the left panel, we also show star-forming galaxies and SMGs at $z=$\ 3--4 from the literature, which 
are introduced in Section~\ref{subsec:comparisonsample}. 
\added{In the right panel, we show 
stacking results at $z\sim3.3$ from the MOSDEF survey \citep{sanders20}. 
We use the line ratios given in \citet{sanders20} and the same metallicity 
calibration method as shown in Section~\ref{subsec:ism}.}
Our targets 
distribute around the star-forming main sequence and the mass-metallicity relation, 
and thus, are not biased in terms of the star-forming activity and gas-phase metallicity. 
The stacked sample is also close to the star-forming main sequence 
and the mass--metallicity relation, 
indicating that the stacked sample has a typical SFR and gas-phase metallicity 
for its stellar mass.

\subsection{Gas mass}\label{subsec:coldgasestimate}

We converted the dust masses from {\sc magphys} to gas masses 
with the relation between the gas-phase metallicity and gas-to-dust mass ratio. 
We use the relation shown in \citet{magdis12} as follows:

\begin{equation}
    {\rm log}(\delta_{\rm gdr}) = (10.54 \pm 1.0) - (0.99 \pm 0.12) \times (\rm 12+log(O/H)), 
\end{equation}

\noindent
which is based on the relation of \citet{leroy11} and 
uses the metallicity estimated with the [{\sc Nii}]/H$\alpha$ ratio 
of \citet{PP04}.
Note that the dust mass estimation in \citet{leroy11} and \citet{magdis12} is based on the 
\citet{draineli07} models. 
The scatter of this relation is $0.15$~dex \citep{magdis12}. 

We need to convert the gas-phase metallicity in Table~\ref{tab:basicquantity}  
to that based on the \citet{PP04} calibration. 
We estimated [{\sc Nii}]/H$\alpha$ ratios using the relation between 
$\rm 12+log(O/H)$ and [{\sc Nii}]/H$\alpha$ of \citet{curti16}, 
and then, converted the estimated [{\sc Nii}]/H$\alpha$ ratios 
to the gas-phase metallicities using the \citet{PP04} calibration. 
\added{The empirical relation between $\rm 12+log(O/H)$ and [{\sc Nii}]/H$\alpha$ 
has a scatter of 0.1~dex along the metallicity direction \citep{curti16}. 
This scatter causes $\sim0.19$~dex uncertainty on average  
on the estimated [{\sc Nii}]/H$\alpha$ ratios for our sample. 
Given that the \citet{PP04} calibration has a scatter of $0.18$~dex, 
the converted gas-phase metallicities have a typical uncertainty 
of $0.26$~dex.}

Then, we estimated gas masses as follows: 

\begin{equation}
   {\rm M_{\rm gas}} = {\rm M_{\rm dust}} \times \delta_{\rm gdr},
   \label{eq.dusttogas}
\end{equation}

\noindent
where $\rm M_{\rm gas}$ includes both molecular and atomic hydrogen. 
We multiply our dust masses by a factor of two when converting them to gas masses 
with Eq.~(\ref{eq.dusttogas}) to correct for the systematic difference of the dust mass estimation  
between {\sc magphys} and \citet{draineli07} models (Section~\ref{subsec:magphys}). 
The estimated gas masses and limits of the individual sources and the stacked sample are summarized 
in Table~\ref{tab:basicquantity}.

Given 
$\sim 0.30$~dex uncertainty coming from the assumed $\kappa_{\rm abs}$ \citep{hunt19}, 
\added{$\sim 0.26$~dex uncertainty on the converted gas-phase metallicities,} 
and $\sim0.15$~dex scatter of the relation between 
the gas-phase metallicity and gas-to-dust mass ratio \citep{magdis12}, 
the systematic uncertainty of our gas mass estimate is roughly 
\added{$0.42$~dex}.

\subsection{Comparison sample from the literature}\label{subsec:comparisonsample}

We next introduce samples from the literature to which we compare our data in Section~\ref{sec:results}. 
Since different works use different approaches to estimate 
dust and/or gas masses, 
these comparisons must be interpreted with care. 
Please refer to the papers cited below for more details  
about the samples selection, observations, and methods used to estimate 
dust and/or gas masses.

\begin{itemize}
\setlength{\leftskip}{-0.5cm} 
\setlength{\itemsep}{-0.1cm}
\item \citet{magdis17} investigated the dust and gas masses 
of two massive Lyman Break Galaxies (LBGs) at $z\sim3$. 
The dust masses are estimated with the \citet{draineli07} models. 
They used several independent methods to estimate 
the gas masses, 
namely, CO(3--2) line, dust mass from the IR SED, 
and the empirical relation of \citet{groves15}. 

\item \citet{wiklind19} targeted star-forming galaxies at $z\sim3$. 
They used the empirical relation of \citet{scoville16} to 
estimate molecular gas masses. 
We show 11 galaxies with the individual molecular gas estimates. 

\item ASPECS: We extract two galaxies 
at $z\sim3.6$ from the ASPECS $1.2$~mm continuum source catalog \citep{aravena20}. 
The dust masses are estimated from SED fitting with {\sc magphys}. 
We use the gas masses estimated with the dust mass and the fixed gas-to-dust mass ratio of 200 
in their catalog. 

\item \citet{cassata20} observed CO emission lines for massive LBGs at $z\sim$ 3--4. 
They used the CO(5--4) emission lines to estimate molecular gas masses. 

\item AS2UDS is an ALMA survey targeting 700 SMGs \citep{dudzeviciute20}. 
Here we show the AS2UDS galaxies at $z=$\ 3--4. 
\citet{dudzeviciute20} estimated the dust masses of the AS2UDS galaxies 
from the SED fitting with {\sc magphys}. 
They converted the dust mass to the gas mass 
assuming the fixed gas-to-dust mass ratio of 100.

\item \citet{tan14} investigated the dust masses of three SMGs at $z=4.05$
with the multi-band photometry in the IR regime. 
They used the \citet{draineli07} dust models 
to estimate the dust masses. 

\end{itemize}

We also introduce the following galaxy samples at lower redshifts,  
which have individual measurements of gas mass and gas-phase metallicity.

\begin{itemize}
  \setlength{\leftskip}{-0.5cm}  
  \setlength{\itemsep}{-0.1cm}
\item \citet{saintonge13} investigated the dust and 
molecular gas masses 
of 17 gravitationally lensed star-forming galaxies at $z\sim$\ 1.4--3. 
They used the \citet{draineli07} models to estimate dust masses.
Molecular gas masses are estimated from the CO(3--2) lines. 
We show four galaxies at $z\sim$ 2--3 in Section~\ref{subsec:msmdust} and \ref{subsec:fgas_OH_comparison}.

\item \citet{seko16_alma} investigated the dust and molecular gas masses 
of star-forming galaxies at $z\sim1.4$. 
They converted the dust continuum flux to a dust mass 
assuming modified black body with fixed $T_{\rm dust}=30$~K and $\beta=1.5$. 
They used the CO(5--4) line to estimate the molecular gas masses.

\item xCOLD-GASS is a CO(1--0) line survey for local SDSS galaxies. 
We use the public catalog in \citet{saintonge17} 
and combine it with the catalog of the xGASS project 
\citep{catinella18} to estimate the total gas masses. 
\added{The stellar masses of the xCOLD-GASS galaxies are in the range of 
$\rm log(M_*/M_\odot)=$ 9.1--11.2}.

\item ALLSMOG is a CO(2--1) line survey for local SDSS galaxies \citep{bothwell14}. 
Most of the ALLSMOG galaxies have the measurements of the atomic hydrogen gas by different studies 
(please see \citet{cicone17} for more details). 
\added{The stellar mass range of the ALLSMOG galaxies 
is $\rm log(M_*/M_\odot)=$ 9.3--10.0}. 

\end{itemize}

\section{Results and Discussion} \label{sec:results}

\subsection{Dust mass and its metallicity dependence}\label{subsec:msmdust}

\begin{figure*}[tb]
    \begin{minipage}[cbt]{0.59\textwidth}
        \centering\includegraphics[width=0.73\textwidth]{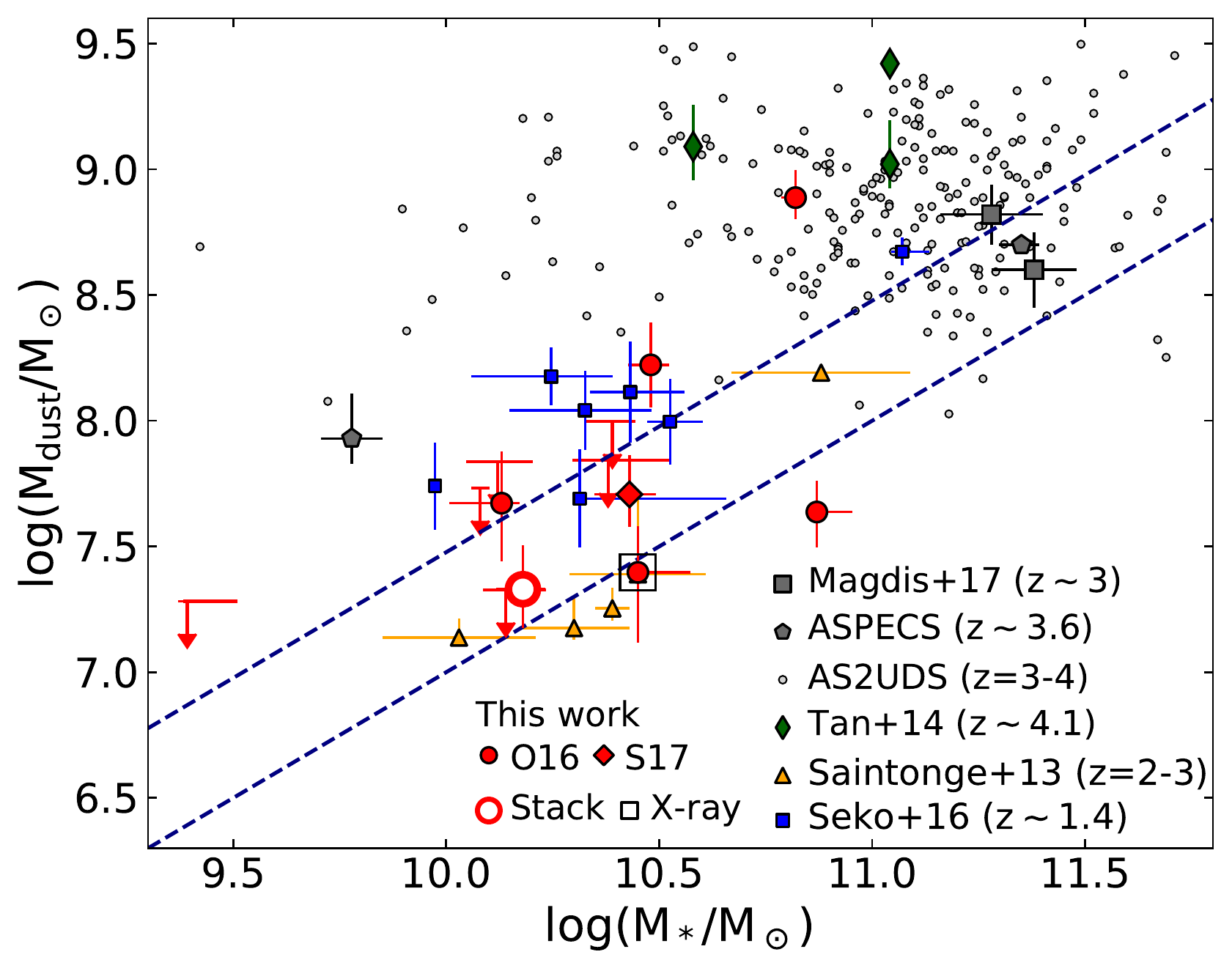}
    \end{minipage}
    \begin{minipage}[cbt]{0.39\textwidth}
        \centering\includegraphics[width=0.92\textwidth]{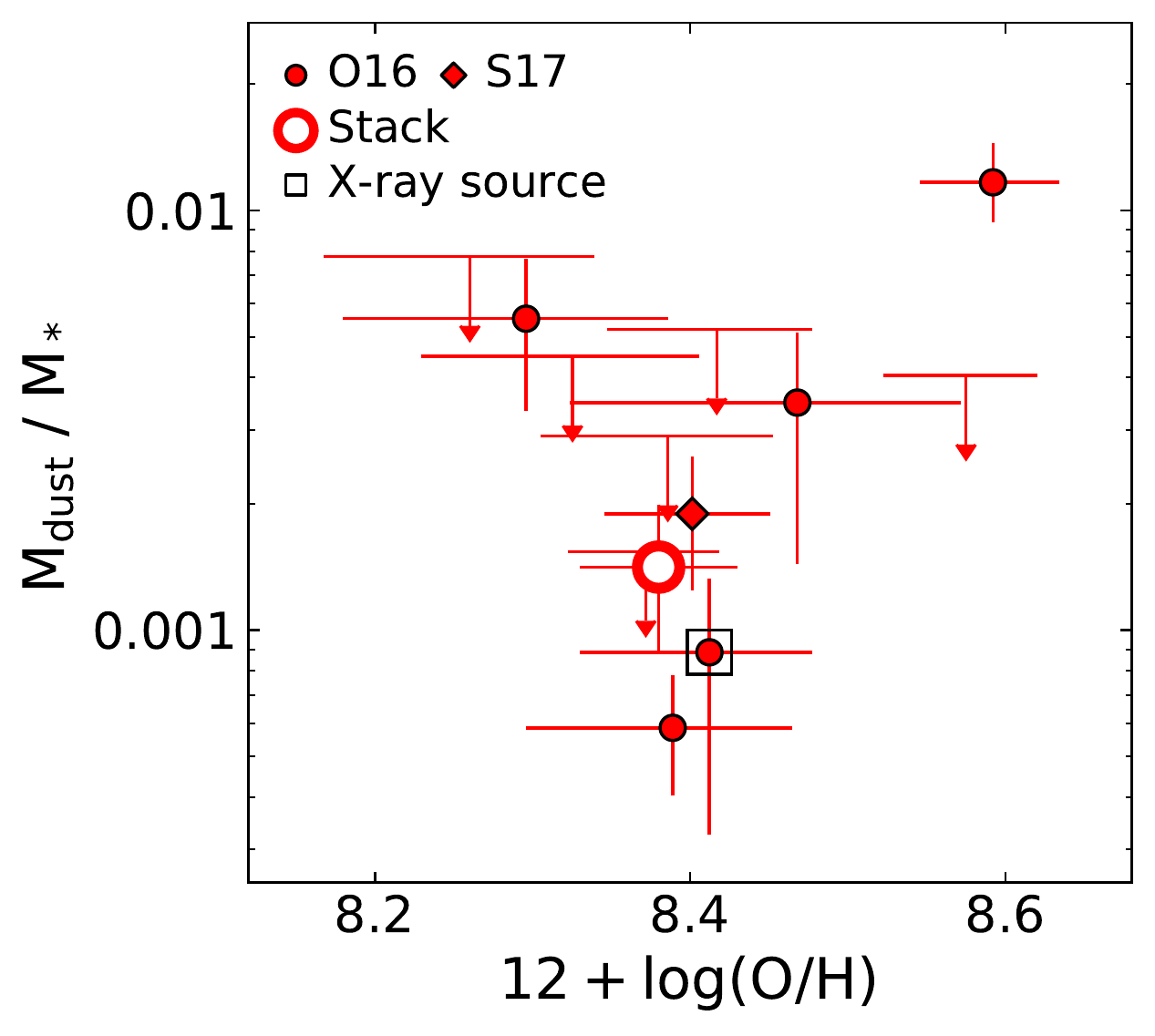}
    \end{minipage}
    \caption{(Left) Relation between stellar mass and dust mass 
    for the star-forming galaxies at $z\sim3.3$ together with galaxies at $z\sim$\ \added{1.4--4} 
    from the literature. 
    The dashed lines correspond to constant dust-to-stellar mass ratios of 
    $\rm M_{dust}/M_* = 1\times10^{-3}$ and $3\times10^{-3}$. 
    We show the galaxy samples of \citet{tan14}, \citet{magdis17}, \added{and \citet{saintonge13}} after dividing their dust masses 
    by a factor of two \added{to correct for the difference of the assumed $\kappa_{\rm abs}$}. 
    The star-forming galaxies at $z\sim3.3$ have similar dust-to-stellar mass ratios as more massive 
    star-forming galaxies from \citet{magdis17} and \added{ASPECS} \citep{aravena20}.  
    (Right) Relation between the gas-phase metallicity and dust-to-stellar mass ratio 
    for the star-forming galaxies at $z\sim3.3$.
    We find no clear correlation between the gas-phase metallicity and dust-to-stellar mass ratio \added{among our sample}.
    }
    \label{fig:ms-mdust}
\end{figure*}

%% Dust mass comparison 
The dust masses of the star-forming galaxies at $z\sim3.3$ 
are estimated to be $\rm log(M_{dust}/M_\odot)\sim$ 7.4--8.9 
(Table~\ref{tab:basicquantity}). 
The dust mass of the stacked sample is $\rm log(M_{dust}/M_\odot)=7.33^{+0.17}_{-0.15}$. 
The left panel of Figure~\ref{fig:ms-mdust} shows the comparison of 
dust masses between the star-forming galaxies at $z\sim3.3$ and 
the galaxies at $z\sim$\ \added{1.4--4} in the literature (Section~\ref{subsec:comparisonsample}). 
As mentioned in Section~\ref{subsec:comparisonsample}, 
\citet{tan14}, \citet{magdis17}, and \added{\citet{saintonge13}} estimated dust masses with the 
\citet{draineli07} models. 
To correct for the systematic difference of the dust mass estimate 
between {\sc magphys} and \citet{draineli07} models, 
the dust masses of the galaxies in these studies %\citet{tan14} and \citet{magdis17} 
are divided by a factor of two in the left panel of Figure~\ref{fig:ms-mdust}.

The right panel of Figure~\ref{fig:ms-mdust} shows the relation between 
the gas-phase metallicity and dust-to-stellar mass ratio for the star-forming galaxies 
at $z\sim3.3$. 
Given that dust is produced from metals, 
we would expect galaxies with higher metallicities to have larger dust masses at a given stellar mass. 
We find no statistically significant trend between the dust-to-stellar mass ratio and gas-phase metallicity 
\added{among our sample}.

The brightest source at 1.3~mm among our sample, 208681 
(Table~\ref{table:obssummary} and Figure~\ref{fig:ms-mdust}), 
has a dust mass of $\rm log(M_{dust}/M_\odot)=8.89_{-0.08}^{+0.11}$, which 
is comparable to those of SMGs at $z\sim$\ 3--4 \citep{tan14,dudzeviciute20}. 
This source can be classified as a SMG 
in terms of its dust content. 
The other five galaxies and the stacked sample have $\sim1$~dex lower dust masses 
than the SMGs at $z\sim$\ 3--4 with similar stellar masses. 
This brightest source, 208681, is also the most metal-rich galaxy with $\rm 12+log(O/H)=8.59_{-0.05}^{+0.04}$ 
among our sample 
\added{and appears to distribute apart from the other galaxies 
in the right panel of Figure~\ref{fig:ms-mdust}. 
This may suggest that SMGs have a different relation 
between the gas-phase metallicity and dust-to-stellar mass ratio 
from UV/optical-selected star-forming galaxies.}

The star-forming galaxies at $z\sim$\ 3--4 except for the SMGs show 
a positive correlation between the stellar mass and dust mass. 
The dust-to-stellar mass ratio takes value roughly between $1\times10^{-3}$ and $5\times 10^{-3}$ 
(median value is $2\times10^{-3}$)
in the stellar mass range of $\rm log(M_*/M_\odot)\sim$\ 10.1--11.4.

When comparing 
with star-forming galaxies 
at $z\sim1.4$ \citep{seko16_alma} and $z\sim$~2--3  
\citep{saintonge13}, 
star-forming galaxies at lower redshifts have 
similar dust-to-stellar mass ratios ($1\times10^{-3}$ -- $5\times10^{-3}$) 
as our galaxies at $z\sim3.3$ with similar stellar masses. 
Although a fair comparison is difficult due to different sample selection 
among the three studies, 
the evolution of the dust-to-stellar mass ratios 
seems to be mild since $z\sim1.4$ to 3.3 as shown in \citet{bethermin15}.

\subsection{Gas properties of star-forming galaxies at $z$=3--4}\label{subsec:fgastdep}

\begin{figure*}[t]
\begin{minipage}[cbt]{0.49\textwidth}
    \centering
    \includegraphics[width=0.8\textwidth]{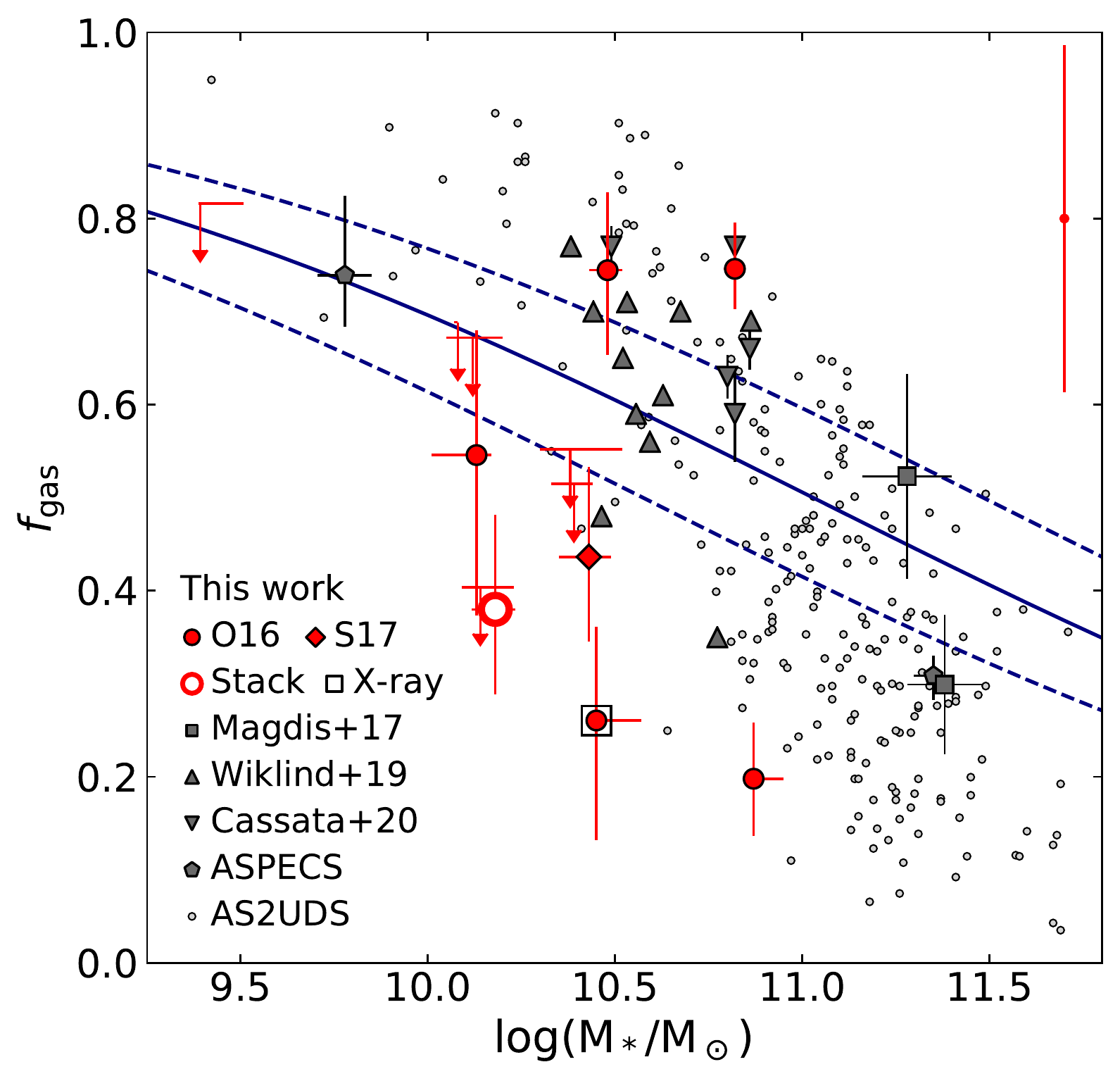}
\end{minipage}
\begin{minipage}[cbt]{0.49\textwidth}
    \centering
    \includegraphics[width=0.8\textwidth]{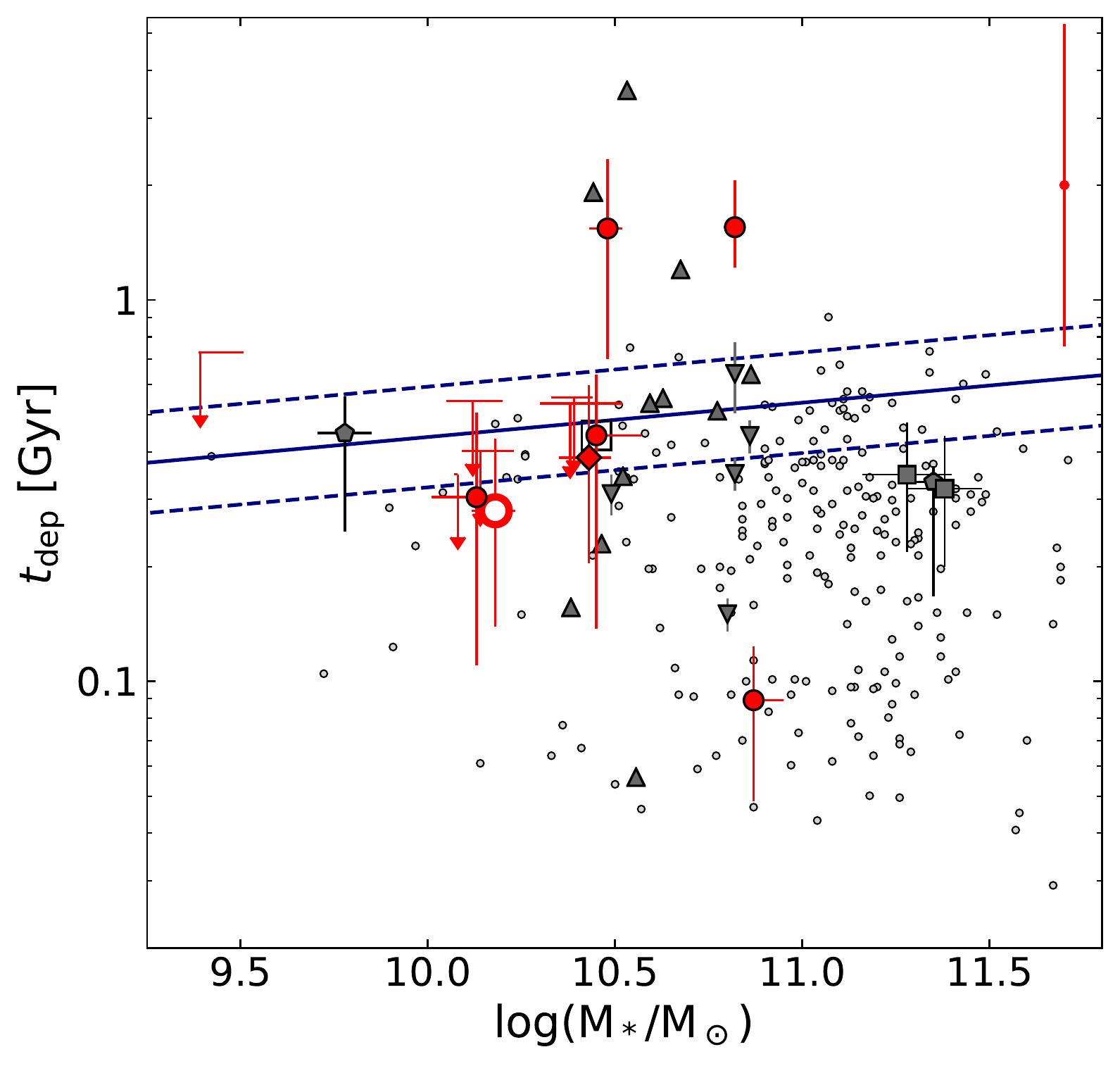}
\end{minipage}
    \caption{Stellar mass versus gas mass fraction (left) and gas depletion timescale (right) diagram 
    for star-forming galaxies at $z\sim$ 3--4. 
    We show our sample at $z\sim3.3$  
    together with galaxies at $z\sim$ 3--4 
    from the literature \citep{magdis17,wiklind19,aravena20,cassata20,dudzeviciute20}. 
    The solid line in each panel represents the scaling relation 
    for star-forming galaxies on the main sequence at $z\sim3.3$ from \citet{tacconi18}. 
    The dashed lines correspond to the cases that galaxies are $0.3\ {\rm dex}$ above/below 
    the main sequence. 
    \added{The vertical line in the top right corner on each panel 
    represents an additional $\pm 1\sigma$ error for our sample coming from the systematic uncertainty on $\rm M_{gas}$ (Section~\ref{subsec:coldgasestimate}).}
    Contrary to the tight distribution of the galaxies at $z\sim3.3$ around the main sequence, 
    the derived gas mass fractions and gas depletion timescales show a large scatter at a fixed stellar mass. 
    Gas properties of star-forming galaxies may have a larger intrinsic scatter than expected from the scaling relation. 
    }
    \label{fig:fgas_tdep}
\end{figure*}

Figure~\ref{fig:fgas_tdep} shows the gas mass fraction, $f_{\rm gas} = {\rm M_{gas}/(M_{gas}+M_*)}$, 
and gas depletion timescale, $t_{\rm dep} = {\rm M_{gas}/SFR}$, as a function of stellar mass 
for the star-forming galaxies at $z\sim3.3$. 
Our estimated gas mass fractions are 0.20--0.75 
and the gas depletion timescales are 0.09--1.55~Gyr. 
The typical uncertainties of $f_{\rm gas}$ and $t_{\rm dep}$ 
are $\pm0.09$ and $\pm0.22$~dex, respectively. 
\added{Given that the gas masses have the systematic uncertainty of 
$\sim0.42$~dex (Section~\ref{subsec:coldgasestimate}),
$f_{\rm gas}$ and $t_{\rm dep}$ have 
an additional error of $\sim0.19$ and 0.42~dex ($1\sigma$), respectively.}
As for the stacked sample, 
the gas mass fraction and gas depletion timescale 
are estimated to be 
$f_{\rm gas}=0.38_{-0.09}^{+0.10}$ and $t_{\rm dep}=0.28_{-0.14}^{+0.15}$~Gyr, respectively.

We show star-forming galaxies and SMGs 
at $z\sim$\ 3--4 from the literature (Section~\ref{subsec:comparisonsample}) in Figure~\ref{fig:fgas_tdep}. 
The solid line in each panel represents the scaling relation  
for galaxies on the star-forming main sequence at $z\sim3.3$ 
from \citet{tacconi18}. 
The dashed lines correspond to the case when galaxies are at $\pm 0.3$~dex 
from the star-forming main sequence. 
Our sample including the stacking result reaches down to 
$f_{\rm gas}\sim$ 0.2--0.3, which is lower by a factor of $\gtrsim2$ than the scaling relation. 
We also find a scatter of $\gtrsim1$~dex for the gas depletion timescale 
at a fixed stellar mass.
Such a large scatter of the gas properties 
is also seen in the samples of \citet{wiklind19} and \citet{dudzeviciute20}.

It has been reported that the gas mass fraction and depletion timescale 
of the main sequence galaxies 
gradually change depending on the deviation from the star-forming main sequence 
($\Delta_{\rm MS}$; e.g., \citealt{saintonge12,sargent14,tacconi18}). 
We checked how the offset of the gas mass fraction or depletion timescale 
from the scaling relation for galaxies on the main sequence 
changes depending on $\rm \Delta_{MS}$. 
We find a trend consistent with \citet{tacconi18} when combining our sample with 
the samples of the literature \citep{magdis17,wiklind19,cassata20,aravena20}. 
However, 
at a fixed $\Delta_{\rm MS}$, 
our sample shows a large scatter of the gas mass fraction and depletion timescale. 
The observed scatter of the gas mass fraction and depletion timescale in our sample 
cannot be explained by $\Delta_{\rm MS}$ alone. 
These results suggest that the fundamental gas properties of galaxies 
have a large diversity even when they have similar stellar masses and SFRs \citep{elbaz18}. 
Given that the scaling relations are possibly biased toward dusty and gas-rich galaxies 
especially at higher redshifts, 
the scaling relations may not be representative of the majority of the galaxy populations 
at $z\sim$ 3--4. 

\added{Given that we use the metallicities to derive the gas properties, 
the observed trends as a function of stellar mass in Figure~\ref{fig:fgas_tdep} 
may be partly caused by the mass--metallicity relation. 
However, the distribution of our sample in Figure~\ref{fig:fgas_tdep} does not change significantly 
when assuming a constant gas-to-dust mass ratio. 
This means that our results are not affected 
by the fact that we use the gas-phase metallicities to derive the gas properties.}

\subsection{Gas mass fraction versus physical conditions of the ionized gas}\label{subsec:fgas_OH}

\begin{figure}
    \centering
    \includegraphics[width=0.4\textwidth]{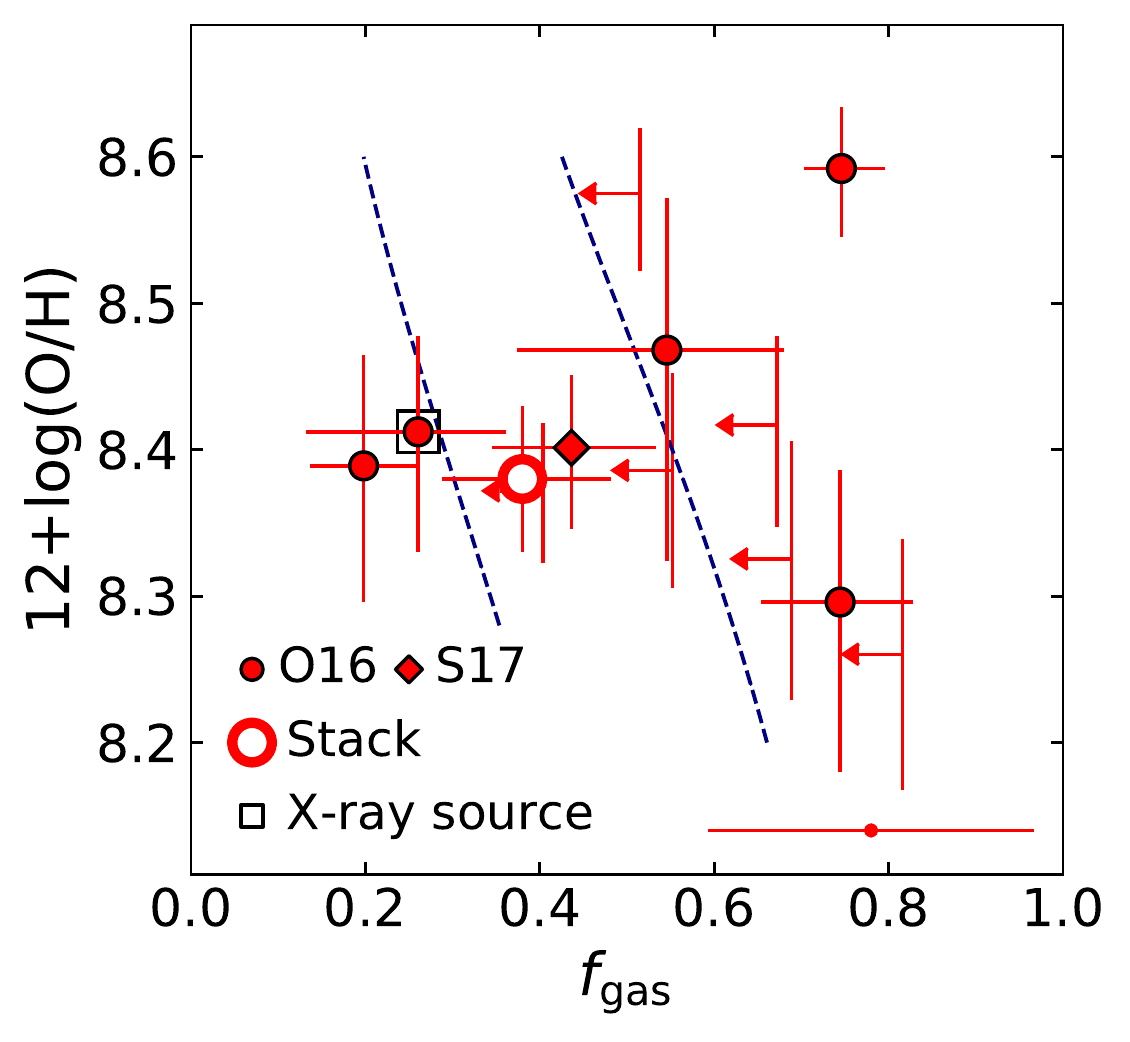}
    \caption{
    Gas mass fraction versus gas-phase metallicity 
for our galaxy sample at $z\sim3.3$. 
\added{The horizontal bar in the bottom right corner 
shows an additional $\pm 1\sigma$ error on the gas mass fraction 
coming from the systematic uncertainty on $\rm M_{gas}$. 
The dashed lines show how the two quantities depend on each other 
when fixing the dust-to-stellar mass ratio at $1\times10^{-3}$ and $3\times10^{-3}$.}
We find no statistically significant correlation 
between the gas mass fraction and gas-phase metallicity among our sample. 
    }
    \label{fig:fgas_OH_z3}
\end{figure}

\added{We investigate the relation between the gas mass fraction 
and the physical conditions of the ionized gas, namely, 
gas-phase metallicity and ionization parameter (Section~\ref{subsec:ism}), 
for our galaxy sample at $z\sim3.3$.
Figure~\ref{fig:fgas_OH_z3} shows the comparison between the gas mass 
fraction and gas-phase metallicity.}
We note that the gas mass fraction and gas-phase metallicity 
(and also ionization parameter) are not independent 
\added{as mentioned in the previous section. 
In Figure~\ref{fig:fgas_OH_z3}, we show how the two quantities 
depend on each other when fixing the dust-to-stellar mass ratio 
with dashed lines. 
}

Given that the abundance of oxygen with respect to hydrogen 
changes depending on the amount of the hydrogen gas in galaxies, 
the gas-phase metallicity, 12+log(O/H), would be expected to decrease 
as the gas mass fraction increases 
\citep[e.g.,][]{bothwell13,zahid_apj791,bothwell16_aa,seko16_alma}. 
However, we find no statistically significant correlation between the gas mass fraction 
and gas-phase metallicity 
for the star-forming galaxies at $z\sim3.3$, 
which is the same as the result obtained from the comparison between the gas-phase metallicity 
and dust-to-stellar mass ratio (Figure~\ref{fig:ms-mdust}).

We find no clear correlation between the gas mass fraction and ionization parameter as well. 
According to \citet{kashino19}, 
the gas mass fraction and ionization parameter are related indirectly via 
three parameters, 
namely, specific SFR (sSFR), gas-phase metallicity, and electron density. 
When the gas-phase metallicity increases or sSFR decreases, both the gas mass fraction and ionization parameter decreases. 
When the electron density increases, the gas mass fraction increases but the ionization parameter decreases \citep{kashino19}. 
Because the gas mass fraction and ionization parameter depend on the three parameters in a different way, 
how the gas mass fraction correlates with the ionization parameter is not straightforward.
We would need to fix some of the parameters to investigate the trend between the two quantities.

A lack of a clear correlation between gas mass fractions 
and the physical conditions of the ionized gas 
may reflect stochastic star-formation histories 
for star-forming galaxies at high redshifts. 
Star formation in galaxies at higher redshifts 
are suggested to be burstier than local galaxies \citep{yicheng16,faucher-giguere18,tacchella20}. 
When the star-forming activity in galaxies changes on a short timescale, 
it becomes more difficult to identify a global trend between the physical quantities. 

Note that our sample size may be too small to find 
any correlation.  
We need a larger sample of galaxies covering a wider range of stellar mass 
to confirm whether 
the gas mass fraction correlates with gas-phase metallicity and ionization parameter.

\subsection{Comparison of galaxies at $z=$ 0--3.3 on the $f_{gas}$ versus 12+log(O/H) diagram}\label{subsec:fgas_OH_comparison}

\begin{figure*}[tb]
    \centering
    \includegraphics[width=0.5\textwidth]{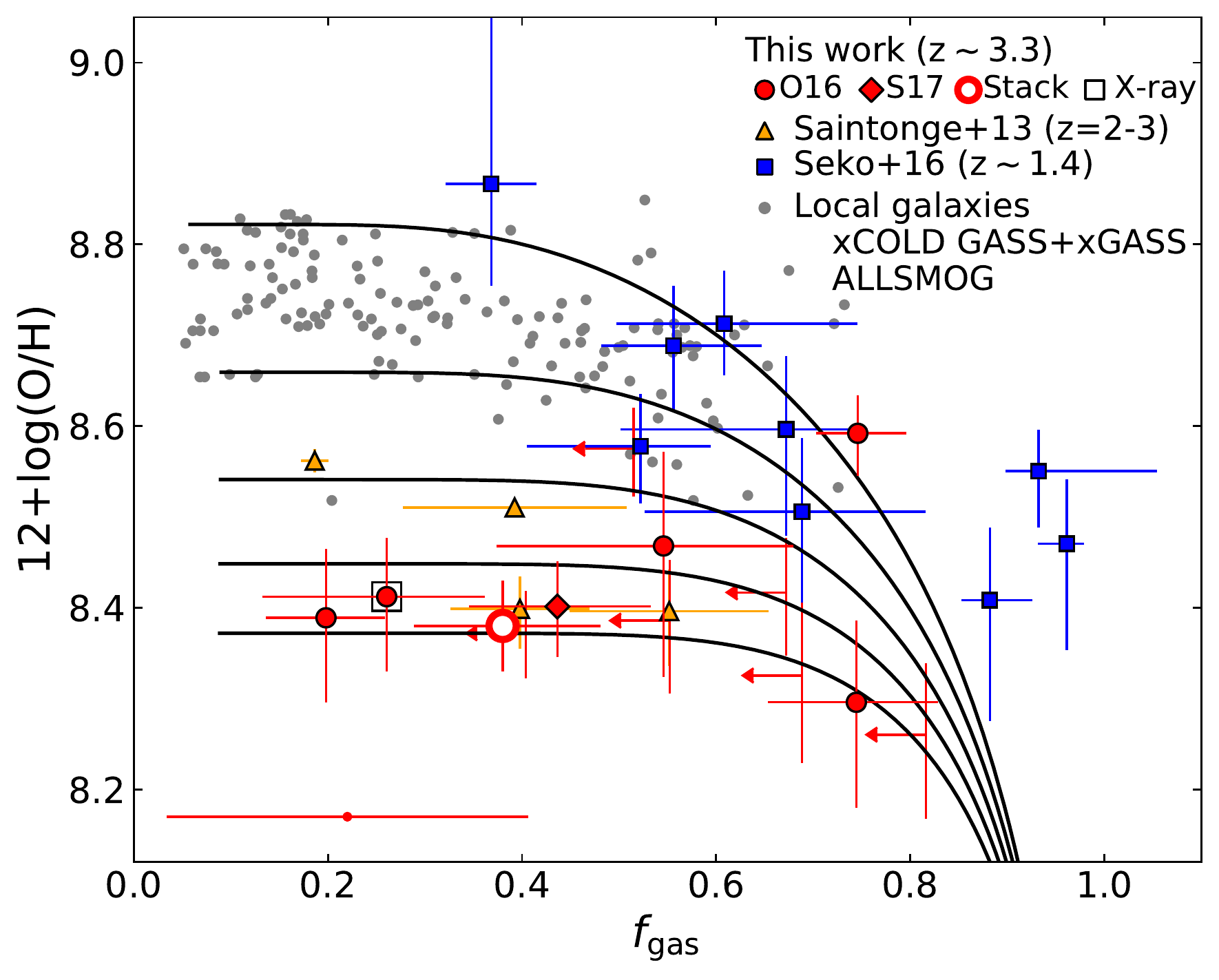}
    \caption{Relation between gas mass fraction and gas-phase metallicity 
    for star-forming galaxies from $z=0$ to $z\sim3.3$.
    \added{The horizontal bar in the left bottom corner represents 
    an additional $\pm 1\sigma$ error on the gas mass fraction 
coming from the systematic uncertainty on $\rm M_{gas}$ for our sample.}
    Only the molecular gas components are considered for the galaxies in \citet{seko16_alma} 
    and \citet{saintonge13}. 
    The star-forming galaxies at $z\gtrsim2$ show an offset 
    toward the lower gas-phase metallicity from the distribution of the local galaxies. 
    The black lines show the model tracks from the gas regulator model of \citet{pengmaiolino14} 
    assuming different mass-loading factors between $\lambda=$ 0.5 and 2.5. 
    The distribution of the star-forming galaxies at $z\sim3.3$ on this diagram 
    can be broadly explained with the model tracks with $\lambda \sim$ 2--2.5, 
    suggesting the redshift evolution of the mass-loading factor for star-forming galaxies. 
    }
    \label{fig:fgas_OH_lowz}
\end{figure*}

Figure~\ref{fig:fgas_OH_lowz}  
shows the star-forming galaxies from $z=0$ to 3.3 (Section~\ref{subsec:comparisonsample}) 
on the gas mass fraction versus metallicity diagram.

In the literature \citep{saintonge13,seko16_alma,saintonge17,cicone17},  
the gas-phase metallicities are estimated with the [{\sc Nii}]/H$\alpha$ ratios.
In order to compare with the gas-phase metallicities of our sample, which are 
estimated based on [{\sc Oiii}], H$\beta$, and [{\sc Oii}] lines \citep{curti16}, 
we convert the given [{\sc Nii}]/H$\alpha$ ratios in the previous studies 
to the gas-phase metallicities using the empirical relation between $\rm 12+log(O/H)$ 
and [{\sc Nii}]/H$\alpha$ 
of \citet{curti16}.

The gas mass fraction of the galaxies 
at $z\sim0$ and $3.3$ 
is the total (molecular$+$atomic) 
gas mass fraction 
to compare with a gas regulator model, in which 
the atomic and molecular hydrogen are indistinguishable, 
in the following sections. 
The gas mass fraction of the galaxies 
in \citet{seko16_alma} and \citet{saintonge13} is the molecular gas mass fraction. 
\color{black}
We expect that the comparison in Figure~\ref{fig:fgas_OH_lowz} 
is not significantly affected by the fact that 
we do not include the atomic gas for the two samples. 
The fraction of the molecular gas is suggested to increase with increasing redshifts 
because of the higher surface density of galaxies at higher redshifts 
\citep[e.g.,][]{popping15}.  
\citet{popping15} suggest the fraction of the molecular gas 
in the total gas is $\sim$\ 0.6--0.8 at $z\sim$\ 1.5--3.0 
based on their simulations.

Focusing on the local galaxies in Figure~\ref{fig:fgas_OH_lowz}, 
the gas-phase metallicity gradually decreases with increasing gas mass fraction 
as shown in \citet{bothwell13} and \citet{hunt15}. 
Such a gradual decrease of the gas-phase metallicity 
with increasing gas mass fraction 
indicates that 
we need to cover a wide range of gas mass fraction 
to identify the correlation between the two quantities.

Whereas the star-forming galaxies at $z\sim1.4$ from \citet{seko16_alma} 
appear to be located at the gas-rich end of the distribution of the local star-forming galaxies, 
the galaxies at $z\gtrsim2$ from this study and \citet{saintonge13} 
show an offset toward the lower gas-phase metallicities ($\sim0.2$~dex)
with respect to the distribution of the local galaxies.  
This result suggests that 
star-forming galaxies at $z\gtrsim2$ are less chemically enriched 
than those at $z=0$ and even at $z\sim1.4$ with similar gas mass fractions.

The molecular gas mass is estimated with the CO lines in the literature (Section~\ref{subsec:comparisonsample}). 
Although the systematic difference caused by using different methods to estimate gas mass 
could change the relative distribution of the galaxies in the horizontal direction, 
it cannot explain the offset of the galaxies at $z\sim3.3$ toward the low gas-phase metallicity 
with respect to the local galaxies. 
As for the gas-phase metallicity, 
\citet{curti16} showed 
that the offset of gas-phase metallicities 
calibrated with different line ratios is 0.04~dex on average. 
The systematic uncertainty caused by using different line ratios to calibrate the metallicity 
is also unlikely to affect our results. 

\added{We have a caveat on our gas-phase metallicity measurement 
for the ALMA-detected, dusty star-forming galaxies in our sample. 
\citet{herrera-camus18} reported that gas-phase metallicities 
calibrated with rest-frame optical emission lines tend to be lower 
than those calibrated with FIR fine-structure lines up to by a factor of two 
for local (U)LIRGs. 
The FIR lines are likely to trace the ionized gas in the dense and dusty star-forming regions, which are no longer traced by the optical emission lines, 
and such dense and dusty star-forming regions would be more metal enriched \citep[e.g.,][]{santini10}.  
When this is also the case for our ALMA-detected galaxies at $z\sim3.3$, 
the offset toward the low metallicity 
with respect to the local galaxies in Figure~\ref{fig:fgas_OH_lowz} could be 
explained by the underestimated gas-phase metallicities for our sample. 
However, when comparing the dust extinction values, $\rm A_V$, between 
our sample and local (U)LIRGs in \citet{rupke08}, 
the median $\rm A_V$ of our ALMA-detected galaxies ($\sim0.6$~mag) 
is much smaller than that of local (U)LIRGs ($\sim3.6$~mag). 
This implies that 
the ALMA-detected galaxies in our sample are not as dusty as 
the local (U)LIRGs, and thus, that 
the metallicities calibrated with the optical emission lines 
can be regarded as representative values for our sample at $z\sim3.3$.}

\subsubsection{Comparison with a gas regulator model} \label{subsec:outflowrate}

``Equilibrium'', ``bathtub'' or ``gas regulator'' models   
are used to track the evolution of the fundamental physical quantities of galaxies, 
such as gas mass, SFR, and metallicity, 
by considering gas inflows, outflows, 
star formation, and metal production in galaxies 
\citep[e.g.,][]{finlator08,bouche10,dave12,Dayal13,lilly13,pengmaiolino14,tacchella20}. 
We compare the observational data at $z=$\ 0--3.3 with a gas regulator model by \citet{pengmaiolino14}.

\citet{pengmaiolino14} derived the analytic formula to track the evolution of the physical quantities, 
such as gas mass, SFR, metallicity, and stellar mass. 
The input parameters of this model are 
gas inflow rate ($\Phi$), star formation efficiency ($\varepsilon =$\ SFR/$\rm M_{gas}$), 
mass-loading factor ($\lambda=$\ outflow rate/SFR), and return mass fraction ($R$). 
The gas accretion to the galaxy is assumed to scale with the growth rate of the dark matter halo. 
The dark matter halo growth rate is derived from the cosmological hydrodynamic simulations \citep{faucher-giguere11}. 
The outflow rate is assumed to be proportional to SFR. 
The return mass fraction takes on values $\sim0.2$ to $\sim0.5$ depending on the IMF. 
This model assumes that these input parameters are constant 
with time or change with longer timescales than 
the equilibrium timescale. 
The equilibrium timescale is 
the timescale to reach the equilibrium state, 
where gas acquisition by inflows balance with gas consumption 
by star formation and outflows. 
The equilibrium timescale is expressed as follows: 

\begin{equation}
    \tau_{\rm eq} = \frac{1}{\varepsilon(1-R+\lambda)}. 
    \label{eq:taueq}
\end{equation}

The time evolution of the gas mass fraction and gas-phase metallicity 
is described as follows: 

\begin{equation}
    f_{\rm gas}(t) = \frac{1}{1 + \varepsilon (1-R) \left(\frac{t}{1-e^{-\frac{t}{\tau_{\rm eq}}}} - \tau_{\rm eq} \right)}, 
\end{equation}

\begin{equation}
    Z_{\rm gas}(t) = [Z_0 + y\tau_{\rm eq}\varepsilon(1 - e^{-\frac{t}{\tau_{\rm eq}}})][1-e^{-\frac{t}{\tau_{\rm eq}(1-e^{-t/\tau_{\rm eq}})}}], 
\end{equation}

\noindent
where 
$Z_0$ is the metallicity of the infalling gas, 
and $y$ is the average yield per stellar generation.

In Figure~\ref{fig:fgas_OH_lowz}, 
we show the model tracks obtained from the gas regulator model of \citet{pengmaiolino14}. 
We assume $R=0.4$ (for the Chabrier IMF; \citealt{madau14}) and 
$y = 1.5 Z_\odot$ \citep[e.g.,][]{yabe15_apj}. 
The gas depletion timescale ($1/\varepsilon$) is set to be $t_{\rm dep} = 0.8\ {\rm Gyr}$.  
Note that the normalization of model tracks in Figure~\ref{fig:fgas_OH_lowz} 
does not depend on the absolute value of $t_{\rm dep}$.
We assume five different mass-loading factors between $\lambda=$\ 0.5 and 2.5 (Figure~\ref{fig:fgas_OH_lowz}).

We find that 
the distribution of the star-forming galaxies at $z\sim3.3$ and 
those from \citet{saintonge13} can be broadly explained by the model tracks with the high mass-loading factor of 
$\lambda \sim$\ 2.0--2.5 rather than the lower values such as $\lambda \sim$\ 0.5 or 1. 
We need stronger outflow with larger $\lambda$ to 
achieve the lower gas-phase metallicity for star-forming galaxies at $z\gtrsim2$  
than the local ones with similar gas mass fractions. 
This result may suggest a redshift evolution of the mass-loading factor $\lambda$ from $z=0$ to $3.3$, as we discuss below.

\citet{yabe15_apj} showed the increasing outflow rate normalized by SFR with increasing redshifts up to $z\sim2$ 
by comparing the observational data (stellar mass, gas mass fraction, and gas-phase metallicity) 
with a simple chemical evolution model (see also \citealt{troncoso14}).
Some theoretical studies 
based on analytic models or numerical simulations 
showed the redshift evolution of the mass-loading factor 
\citep[e.g.,][]{barai15,mitra15,muratov15,hayword17}. 
Observationally, \citet{sugahara17} showed a trend 
that the star-forming galaxies at higher redshift (up to $z=2$)
have larger mass-loading factor at a fixed circular velocity. 
Our results obtained from the comparison 
between the observational data and the model tracks 
support the idea that star-forming galaxies at higher redshifts have larger mass-loading factors,  
and thus, more massive outflow.

\subsubsection{Equilibrium timescale}

We estimate the equilibrium timescales (Eq.~(\ref{eq:taueq})) for the star-forming galaxies at $z\sim3.3$ 
with the gas depletion timescale obtained from the observation 
and the mass-loading factor inferred from the comparison with the model tracks 
in Figure~\ref{fig:fgas_OH_lowz}. 
The equilibrium timescales of the ALMA-detected sources are estimated to be 0.03--2.21~Gyr (average value: 0.52~Gyr) 
assuming $R=0.4$ for the Chabrier IMF \citep{madau14}.

According to \citet{pengmaiolino14},  
when the equilibrium timescale is much shorter than the Hubble time, 
galaxies are expected to be in the equilibrium state, 
where gas acquisition by inflows and 
gas consumption by star formation and outflows are balanced. 
On the other hand,
when the equilibrium timescale is comparable to the Hubble time, 
galaxies are considered to have much larger gas reservoir 
and to be out of equilibrium. 

The average equilibrium timescale of the detected sources 
is roughly one order of magnitude shorter than the Hubble time at $z=3.3$ ($2.82$~Gyr). 
However, not
all of the galaxies necessarily start 
forming stars at the beginning of the Universe.
Given that the age of galaxies must be smaller than the Hubble time, 
the equilibrium timescale should probably be compared with the age of the 
galaxies rather than the Hubble time.

We here use the ratio of $\rm M_*/SFR$, which can be  
regarded as the minimum age of a galaxy. 
The star-forming galaxies at $z\sim3.3$ have $\rm M_*/SFR=$ 0.25--1.25~Gyr 
(average: 0.57~Gyr), 
which is closer to the equilibrium timescales 
than the Hubble time. 
Especially, the galaxies with relatively larger gas mass fractions, 
$f_{\rm gas}\sim$\ 0.6--0.8, in our sample 
tend to have the equilibrium timescales comparable to the minimum ages. 
This result may suggest that 
normal star-forming galaxies at $z\sim3$ with relatively large gas mass fractions 
have not yet reached the equilibrium state  
as suggested in \citet{mannucci10}. 

In the future, it will be of interest to study how our results 
are affected by relaxing the assumptions about galaxies being in equilibrium 
and by considering bursty star formation histories \citep{tacchella20}. 
\added{Direct measurements of gas outflow and inflow rates would be 
also important to further investigate whether 
the star-forming galaxies at $z\sim3.3$ are out of equilibrium. 
The spatially resolved emission line maps for the individual galaxies 
will enable us to search for outflow signatures and 
estimate the mass outflow rates \citep[e.g.,][]{genzel11,davies19}. 
Furthermore, 
a simulation study suggests a correlation between metallicity gradients 
and gas accretion rates \citep{collacchioni20}. 
We would be able to investigate the gas inflow rates by obtaining 
metallicity gradients from the spatially resolved emission line maps.}

\section{Summary} \label{sec:summary}
We conducted ALMA Band-6 observations 
of star-forming galaxies at $z\sim3.3$, 
which have measurements of their metallicities based on 
the rest-frame optical spectroscopy. 
Thus we can directly compare the metallicities  
with the dust and inferred gas properties from our ALMA observations 
for star-forming galaxies at $z\sim3.3$. 
We detected the dust continuum emission individually from six out of 12 galaxies.  
We stacked the ALMA maps of the five ALMA non-detected sources with 
$\rm log(M_*/M_\odot)=$\ 10.0--10.4 and obtained a $\sim5\sigma$ detection of this sample.

We estimated dust masses from SED fitting with {\sc magphys} 
including the $1.3$~mm fluxes from ALMA. 
We converted the dust mass to the gas mass 
with a relation between the gas-phase metallicity and gas-to-dust mass ratio. 
With the estimates of dust mass, gas mass, and the physical conditions of the ionized gas, 
we conclude the following:

\begin{itemize}
    \item The median value of the dust-to-stellar mass ratios is 
    $\rm M_{dust}/M_* \sim 3.0 \pm 2.0 \times 10^{-3}$.  
    The dust-to-stellar mass ratio of the stacked sample is $\sim 1.4\pm0.5 \times 10^{-3}$. 
    We find no clear trend between the dust-to-stellar mass ratio and gas-phase metallicity.

    \item \added{The estimated gas mass fractions and gas depletion timescales are $f_{\rm gas}=$ 0.20--0.75 and $t_{\rm dep}=$ 0.09--1.55~Gyr, respectively.}
    The stacked sample shows $f_{\rm gas}=0.38_{-0.09}^{+0.10}$ and $t_{\rm dep}=0.28_{-0.14}^{+0.15}$~Gyr. 
    The gas mass fractions and gas depletion timescales of the galaxies at $z\sim3.3$ show a wider spread 
    at a fixed stellar mass as compared to the scaling relations of galaxies on the  
    main sequence at $z\sim3.3$. 
    Given that most of our galaxies at $z\sim3.3$ distribute around the star-forming main sequence with $\pm0.3$~dex, 
    the large scatter of the gas mass fraction and depletion timescale may suggest a significant diversity 
    of these fundamental properties within the so-called main sequence.

    \item We find no clear correlation between the gas mass fraction and 
    the physical conditions of the ionized gas, namely, 
    gas-phase metallicity and ionization parameter, at $z\sim3.3$. 
    We may require a large sample of galaxies covering a wider range of the physical quantities 
    to confirm whether gas mass fractions correlate with the ionized gas conditions or not.

    \item Comparing star-forming galaxies at different redshifts 
    on the gas mass fraction versus metallicity diagram, 
    we find that the star-forming galaxies at $z\gtrsim2$ 
    show an offset toward lower metallicities 
    as compared to the distribution of local star-forming galaxies, 
    in the sense that star-forming galaxies 
    at $z\gtrsim2$ appear to be more metal-poor 
    than the local galaxies with similar gas mass fractions.

    \item We find that the distribution of star-forming galaxies at $z\sim3.3$ on the 
    gas mass fraction versus gas-phase metallicity diagram can be broadly explained by 
    models assuming higher mass-loading factors in outflows of $\lambda\sim$\ 2.0--2.5 
    from the gas regulator model of \citet{pengmaiolino14}. 
    This result supports the idea that star-forming galaxies at higher redshfits have 
    powerful outflows with higher mass-loading factors.

    \item Comparing the equilibrium timescales \citep{pengmaiolino14} and the minimum ages of the galaxies ($\rm M_*/SFR$), 
    we find that the equilibrium timescale of the relatively gas-rich galaxies ($f_{\rm gas}\sim$~0.7) 
    is comparable to their minimum ages, 
    suggesting that they may be out of equilibrium. 
    
\end{itemize}

It remains unclear whether star-forming galaxies at high redshifts follow the same relation 
between gas-phase metallicity and gas-to-dust mass ratio as local galaxies 
(\citealt{saintonge13,seko16_alma}, but see also \citealt{magdis12,shapley20}). 
Observations of independent gas tracers, 
such as CO or [{\sc CI}] emission lines, will be required to 
investigate the relation between the gas-phase metallicity and gas-to-dust mass ratio at $z>3$.

\added{Another caveat is whether the metallicities derived 
from the rest-frame optical emission lines are applicable to 
dusty star-forming galaxies at high redshifts. 
Metallicity measurements with FIR fine structure lines are required to 
investigate this further.}

High-resolution integral-field-unit (IFU) observation 
with 
{\it the James Webb Space Telescope (JWST)} 
will enable us to investigate the metallicity gradients within 
the individual galaxies and to search for the outflow signatures within them. 
\added{The spatially resolved emission line maps} would be useful 
to investigate the effects of gas inflows and outflows more directly.

\acknowledgments
We thank the anonymous referee for careful reading and comments 
that improved the clarity of this paper. 
TLS would like to thank Ken-ichi Tadaki and Nao Fukagawa for useful comments. 
IRS acknowledges support from STFC (ST/T000244/1).
This paper makes use of the following ALMA data: ADS/JAO.ALMA\#2018.1.00681.S. 
ALMA is a partnership of ESO (representing its member states), NSF (USA) and NINS (Japan), 
together with NRC (Canada), MOST and ASIAA (Taiwan), and KASI (Republic of Korea), in 
cooperation with the Republic of Chile. The Joint ALMA Observatory is operated by 
ESO, AUI/NRAO and NAOJ.
Some of the data presented herein were obtained at the W. M. Keck Observatory, 
which is operated as a scientific partnership among the California Institute of Technology, 
the University of California and the National Aeronautics and Space Administration. 
The Observatory was made possibility by the generous financial support of the W. M. Keck Foundation. 
The authors wish to recognize and acknowledge the very significant cultural role 
and reverence that the summit of Maunakea has always had within the indigenous Hawaiian community. 
We are most fortune to have the opportunity to conduct observations from this mountain. 
Data analyses were in part carried out on the open use data
analysis computer system at the Astronomy Data Center, ADC, of 
the National Astronomical Observatory of Japan (NAOJ).


\vspace{5mm}
\facilities{ALMA, Keck:I (MOSFIRE)}


\software{astropy \citep{astropy:2013,astropy:2018}, 
{\sc casa} \citep{CASA},  
{\sc topcat} \citep{topcat}
          }



\appendix
\section{{\sc magphys} best-fit SEDs}\label{sec:appendix1}

\begin{figure*}
\centering\includegraphics[width=1.0\textwidth]{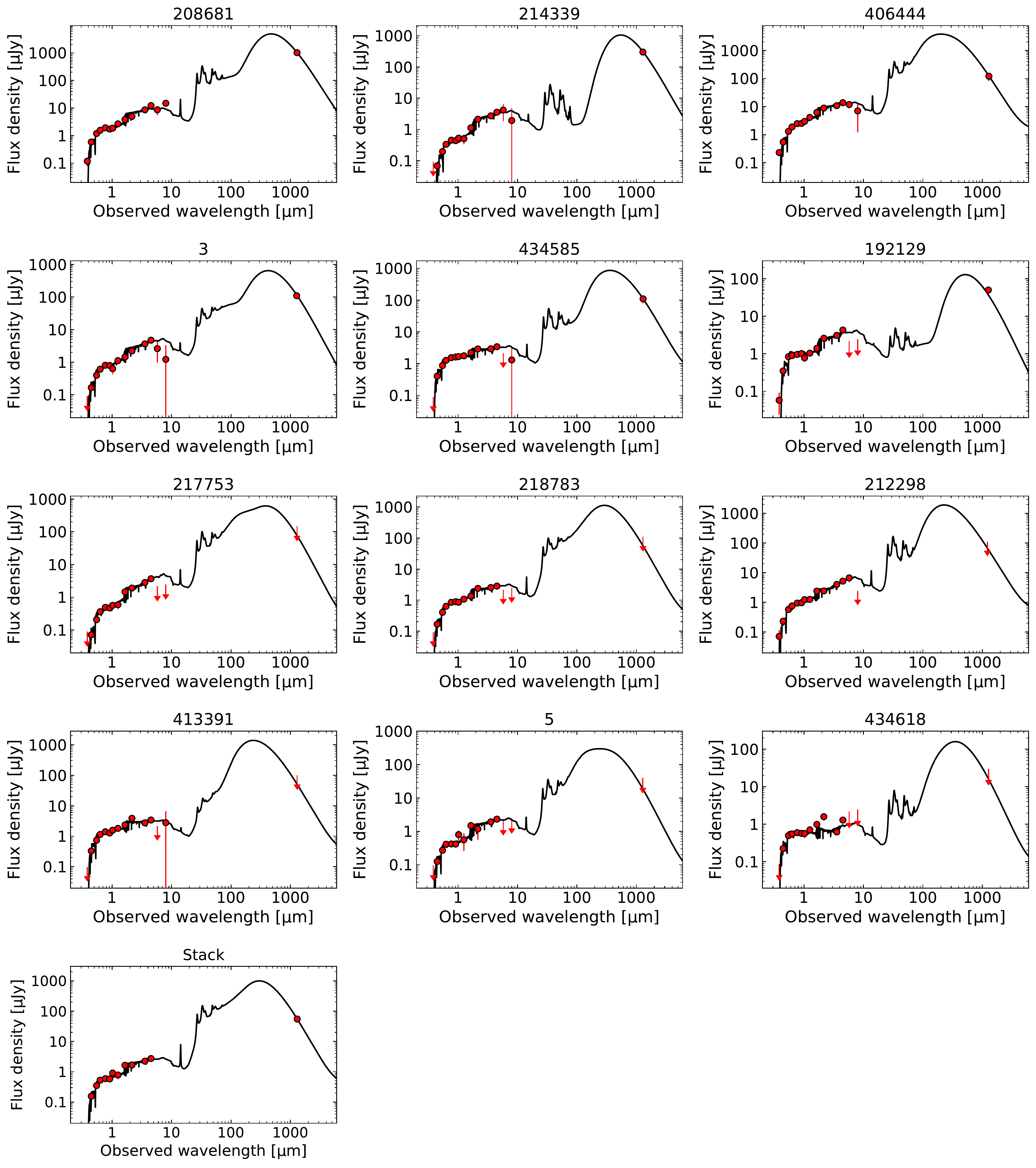}
\caption{Best-fit SEDs (sold line) obtained with {\sc magphys}.
Data points are from the COSMOS2015 catalog \citet{cosmos2015} and the ALMA Band-6 observation in this study. 
Arrows show $3\sigma$ upper limits. 
When fitting the ALMA non-detected sources, 
the 1.3 mm flux and its uncertainty are set to $1.5\sigma \pm 1\sigma$ \citep{dudzeviciute20}. 
}
\label{fig:bestfit-SED}
\end{figure*}

Figure~\ref{fig:bestfit-SED} shows the best-fit SEDs from {\sc magphys} for 12 galaxies observed with ALMA. 
We also show the best-fit SED for the stacked sample including the five ALMA non-detected sources with $\rm log(M_*/M_\odot)=$\ 10.0--10.4 (Section~\ref{subsec:stack}). 

%\bibliography{reference}{}

\begin{thebibliography}{}
\expandafter\ifx\csname natexlab\endcsname\relax\def\natexlab#1{#1}\fi
\providecommand{\url}[1]{\href{#1}{#1}}
\providecommand{\dodoi}[1]{doi:~\href{http://doi.org/#1}{\nolinkurl{#1}}}
\providecommand{\doeprint}[1]{\href{http://ascl.net/#1}{\nolinkurl{http://ascl.net/#1}}}
\providecommand{\doarXiv}[1]{\href{https://arxiv.org/abs/#1}{\nolinkurl{https://arxiv.org/abs/#1}}}

\bibitem[{{Aravena} {et~al.}(2020){Aravena}, {Boogaard},
  {G{\'o}nzalez-L{\'o}pez}, {Decarli}, {Walter}, {Carilli}, {Smail}, {Weiss},
  {Assef}, {Bauer}, {Bouwens}, {Cortes}, {Cox}, {da Cunha}, {Daddi},
  {D{\'\i}az-Santos}, {Inami}, {Ivison}, {Novak}, {Popping}, {Riechers}, {van
  der Werf}, \& {Wagg}}]{aravena20}
{Aravena}, M., {Boogaard}, L., {G{\'o}nzalez-L{\'o}pez}, J., {et~al.} 2020,
  arXiv e-prints, arXiv:2006.04284.
\newblock \doarXiv{2006.04284}

\bibitem[{{Astropy Collaboration} {et~al.}(2013){Astropy Collaboration},
  {Robitaille}, {Tollerud}, {Greenfield}, {Droettboom}, {Bray}, {Aldcroft},
  {Davis}, {Ginsburg}, {Price-Whelan}, {Kerzendorf}, {Conley}, {Crighton},
  {Barbary}, {Muna}, {Ferguson}, {Grollier}, {Parikh}, {Nair}, {Unther},
  {Deil}, {Woillez}, {Conseil}, {Kramer}, {Turner}, {Singer}, {Fox}, {Weaver},
  {Zabalza}, {Edwards}, {Azalee Bostroem}, {Burke}, {Casey}, {Crawford},
  {Dencheva}, {Ely}, {Jenness}, {Labrie}, {Lim}, {Pierfederici}, {Pontzen},
  {Ptak}, {Refsdal}, {Servillat}, \& {Streicher}}]{astropy:2013}
{Astropy Collaboration}, {Robitaille}, T.~P., {Tollerud}, E.~J., {et~al.} 2013,
  \aap, 558, A33, \dodoi{10.1051/0004-6361/201322068}

\bibitem[{{Baldwin} {et~al.}(1981){Baldwin}, {Phillips}, \& {Terlevich}}]{bpt}
{Baldwin}, J.~A., {Phillips}, M.~M., \& {Terlevich}, R. 1981, \pasp, 93, 5,
  \dodoi{10.1086/130766}

\bibitem[{{Barai} {et~al.}(2015){Barai}, {Monaco}, {Murante}, {Ragagnin}, \&
  {Viel}}]{barai15}
{Barai}, P., {Monaco}, P., {Murante}, G., {Ragagnin}, A., \& {Viel}, M. 2015,
  \mnras, 447, 266, \dodoi{10.1093/mnras/stu2340}

\bibitem[{{Battisti} {et~al.}(2019){Battisti}, {da Cunha}, {Grasha}, {Salvato},
  {Daddi}, {Davies}, {Jin}, {Liu}, {Schinnerer}, {Vaccari}, \& {COSMOS
  Collaboration}}]{battisti19}
{Battisti}, A.~J., {da Cunha}, E., {Grasha}, K., {et~al.} 2019, \apj, 882, 61,
  \dodoi{10.3847/1538-4357/ab345d}

\bibitem[{{Best} {et~al.}(2013){Best}, {Smail}, {Sobral}, {Geach}, {Garn},
  {Ivison}, {Kurk}, {Dalton}, {Cirasuolo}, \& {Casali}}]{best13}
{Best}, P., {Smail}, I., {Sobral}, D., {et~al.} 2013, in Astrophysics and Space
  Science Proceedings, Vol.~37, Thirty Years of Astronomical Discovery with
  UKIRT, ed. A.~{Adamson}, J.~{Davies}, \& I.~{Robson}, 235,
  \dodoi{10.1007/978-94-007-7432-2_22}

\bibitem[{{B{\'e}thermin} {et~al.}(2015){B{\'e}thermin}, {Daddi}, {Magdis},
  {Lagos}, {Sargent}, {Albrecht}, {Aussel}, {Bertoldi}, {Buat}, {Galametz},
  {Heinis}, {Ilbert}, {Karim}, {Koekemoer}, {Lacey}, {Le Floc'h}, {Navarrete},
  {Pannella}, {Schreiber}, {Smol{\v{c}}i{\'c}}, {Symeonidis}, \&
  {Viero}}]{bethermin15}
{B{\'e}thermin}, M., {Daddi}, E., {Magdis}, G., {et~al.} 2015, \aap, 573, A113,
  \dodoi{10.1051/0004-6361/201425031}

\bibitem[{{Birkin} {et~al.}(2020){Birkin}, {Weiss}, {Wardlow}, {Smail},
  {Swinbank}, {Dudzevi{\v{c}}i{\={u}}t{\.{e}}}, {An}, {Ao}, {Chapman}, {Chen},
  {da Cunha}, {Dannerbauer}, {Gullberg}, {Hodge}, {Ikarashi}, {Ivison},
  {Matsuda}, {Stach}, {Walter}, {Wang}, \& {van der Werf}}]{birkin20}
{Birkin}, J.~E., {Weiss}, A., {Wardlow}, J.~L., {et~al.} 2020, arXiv e-prints,
  arXiv:2009.03341.
\newblock \doarXiv{2009.03341}

\bibitem[{{Bothwell} {et~al.}(2016{\natexlab{a}}){Bothwell}, {Maiolino},
  {Cicone}, {Peng}, \& {Wagg}}]{bothwell16_aa}
{Bothwell}, M.~S., {Maiolino}, R., {Cicone}, C., {Peng}, Y., \& {Wagg}, J.
  2016{\natexlab{a}}, \aap, 595, A48, \dodoi{10.1051/0004-6361/201527918}

\bibitem[{{Bothwell} {et~al.}(2013{\natexlab{a}}){Bothwell}, {Maiolino},
  {Kennicutt}, {Cresci}, {Mannucci}, {Marconi}, \& {Cicone}}]{bothwell13}
{Bothwell}, M.~S., {Maiolino}, R., {Kennicutt}, R., {et~al.}
  2013{\natexlab{a}}, \mnras, 433, 1425, \dodoi{10.1093/mnras/stt817}

\bibitem[{{Bothwell} {et~al.}(2016{\natexlab{b}}){Bothwell}, {Maiolino},
  {Peng}, {Cicone}, {Griffith}, \& {Wagg}}]{bothwell16_mnras}
{Bothwell}, M.~S., {Maiolino}, R., {Peng}, Y., {et~al.} 2016{\natexlab{b}},
  \mnras, 455, 1156, \dodoi{10.1093/mnras/stv2121}

\bibitem[{{Bothwell} {et~al.}(2013{\natexlab{b}}){Bothwell}, {Smail},
  {Chapman}, {Genzel}, {Ivison}, {Tacconi}, {Alaghband -Zadeh}, {Bertoldi},
  {Blain}, {Casey}, {Cox}, {Greve}, {Lutz}, {Neri}, {Omont}, \&
  {Swinbank}}]{bothwell13_mnras429}
{Bothwell}, M.~S., {Smail}, I., {Chapman}, S.~C., {et~al.} 2013{\natexlab{b}},
  \mnras, 429, 3047, \dodoi{10.1093/mnras/sts562}

\bibitem[{{Bothwell} {et~al.}(2014){Bothwell}, {Wagg}, {Cicone}, {Maiolino},
  {M{\o}ller}, {Aravena}, {De Breuck}, {Peng}, {Espada}, {Hodge},
  {Impellizzeri}, {Mart{\'\i}n}, {Riechers}, \& {Walter}}]{bothwell14}
{Bothwell}, M.~S., {Wagg}, J., {Cicone}, C., {et~al.} 2014, \mnras, 445, 2599,
  \dodoi{10.1093/mnras/stu1936}

\bibitem[{{Bouch{\'e}} {et~al.}(2010){Bouch{\'e}}, {Dekel}, {Genzel}, {Genel},
  {Cresci}, {F{\"o}rster Schreiber}, {Shapiro}, {Davies}, \&
  {Tacconi}}]{bouche10}
{Bouch{\'e}}, N., {Dekel}, A., {Genzel}, R., {et~al.} 2010, \apj, 718, 1001,
  \dodoi{10.1088/0004-637X/718/2/1001}

\bibitem[{{Brown} {et~al.}(2018){Brown}, {Cortese}, {Catinella}, \&
  {Kilborn}}]{brown18}
{Brown}, T., {Cortese}, L., {Catinella}, B., \& {Kilborn}, V. 2018, \mnras,
  473, 1868, \dodoi{10.1093/mnras/stx2452}

\bibitem[{{Bruzual} \& {Charlot}(2003)}]{bc03}
{Bruzual}, G., \& {Charlot}, S. 2003, \mnras, 344, 1000,
  \dodoi{10.1046/j.1365-8711.2003.06897.x}
  
\bibitem[{{Calzetti} {et~al.}(2000){Calzetti}, {Armus}, {Bohlin}, {Kinney},
  {Koornneef}, \& {Storchi-Bergmann}}]{calzetti00}
{Calzetti}, D., {Armus}, L., {Bohlin}, R.~C., {et~al.} 2000, \apj, 533, 682,
  \dodoi{10.1086/308692}

\bibitem[{{Carilli} \& {Walter}(2013)}]{carilliwalter13}
{Carilli}, C.~L., \& {Walter}, F. 2013, \araa, 51, 105,
  \dodoi{10.1146/annurev-astro-082812-140953}

\bibitem[{{Cassata} {et~al.}(2020){Cassata}, {Liu}, {Groves}, {Schinnerer},
  {Ibar}, {Sargent}, {Karim}, {Talia}, {F{\`e}vre}, {Tasca}, {Lemaux},
  {Ribeiro}, {Fiore}, {Romano}, {Mancini}, {Morselli}, {Rodighiero},
  {Rodr{\'\i}guez-Mu{\~n}oz}, {Enia}, \& {Smolcic}}]{cassata20}
{Cassata}, P., {Liu}, D., {Groves}, B., {et~al.} 2020, \apj, 891, 83,
  \dodoi{10.3847/1538-4357/ab7452}

\bibitem[{{Catinella} {et~al.}(2018){Catinella}, {Saintonge}, {Janowiecki},
  {Cortese}, {Dav{\'e}}, {Lemonias}, {Cooper}, {Schiminovich}, {Hummels},
  {Fabello}, {Ger{\'e}b}, {Kilborn}, \& {Wang}}]{catinella18}
{Catinella}, B., {Saintonge}, A., {Janowiecki}, S., {et~al.} 2018, \mnras, 476,
  875, \dodoi{10.1093/mnras/sty089}

\bibitem[{{Chabrier}(2003)}]{chabrier03}
{Chabrier}, G. 2003, \pasp, 115, 763, \dodoi{10.1086/376392}

\bibitem[{{Charlot} \& {Fall}(2000)}]{charlotfall00}
{Charlot}, S., \& {Fall}, S.~M. 2000, \apj, 539, 718, \dodoi{10.1086/309250}

\bibitem[{{Chen} {et~al.}(2017){Chen}, {Hodge}, {Smail}, {Swinbank}, {Walter},
  {Simpson}, {Calistro Rivera}, {Bertoldi}, {Brandt}, {Chapman}, {da Cunha},
  {Dannerbauer}, {De Breuck}, {Harrison}, {Ivison}, {Karim}, {Knudsen},
  {Wardlow}, {Wei{\ss}}, \& {van der Werf}}]{chen17}
{Chen}, C.-C., {Hodge}, J.~A., {Smail}, I., {et~al.} 2017, \apj, 846, 108,
  \dodoi{10.3847/1538-4357/aa863a}


\bibitem[{{Cicone} {et~al.}(2017){Cicone}, {Bothwell}, {Wagg}, {M{\o}ller}, {De
  Breuck}, {Zhang}, {Mart{\'\i}n}, {Maiolino}, {Severgnini}, {Aravena},
  {Belfiore}, {Espada}, {Fl{\"u}tsch}, {Impellizzeri}, {Peng}, {Raj},
  {Ram{\'\i}rez-Olivencia}, {Riechers}, \& {Schawinski}}]{cicone17}
{Cicone}, C., {Bothwell}, M., {Wagg}, J., {et~al.} 2017, \aap, 604, A53,
  \dodoi{10.1051/0004-6361/201730605}

\bibitem[{{Civano} {et~al.}(2012){Civano}, {Elvis}, {Brusa}, {Comastri},
  {Salvato}, {Zamorani}, {Aldcroft}, {Bongiorno}, {Capak}, {Cappelluti},
  {Cisternas}, {Fiore}, {Fruscione}, {Hao}, {Kartaltepe}, {Koekemoer}, {Gilli},
  {Impey}, {Lanzuisi}, {Lusso}, {Mainieri}, {Miyaji}, {Lilly}, {Masters},
  {Puccetti}, {Schawinski}, {Scoville}, {Silverman}, {Trump}, {Urry},
  {Vignali}, \& {Wright}}]{civano12}
{Civano}, F., {Elvis}, M., {Brusa}, M., {et~al.} 2012, \apjs, 201, 30,
  \dodoi{10.1088/0067-0049/201/2/30}

\bibitem[{{Civano} {et~al.}(2016){Civano}, {Marchesi}, {Comastri}, {Urry},
  {Elvis}, {Cappelluti}, {Puccetti}, {Brusa}, {Zamorani}, {Hasinger},
  {Aldcroft}, {Alexander}, {Allevato}, {Brunner}, {Capak}, {Finoguenov},
  {Fiore}, {Fruscione}, {Gilli}, {Glotfelty}, {Griffiths}, {Hao}, {Harrison},
  {Jahnke}, {Kartaltepe}, {Karim}, {LaMassa}, {Lanzuisi}, {Miyaji}, {Ranalli},
  {Salvato}, {Sargent}, {Scoville}, {Schawinski}, {Schinnerer}, {Silverman},
  {Smolcic}, {Stern}, {Toft}, {Trakhenbrot}, {Treister}, \&
  {Vignali}}]{civano16}
{Civano}, F., {Marchesi}, S., {Comastri}, A., {et~al.} 2016, \apj, 819, 62,
  \dodoi{10.3847/0004-637X/819/1/62}

\bibitem[{{Collacchioni} {et~al.}(2020){Collacchioni}, {Lagos}, {Mitchell},
  {Schaye}, {Wisnioski}, {Cora}, \& {Correa}}]{collacchioni20}
{Collacchioni}, F., {Lagos}, C. D.~P., {Mitchell}, P.~D., {et~al.} 2020,
  \mnras, 495, 2827, \dodoi{10.1093/mnras/staa1334}

\bibitem[{{Cresci} {et~al.}(2010){Cresci}, {Mannucci}, {Maiolino}, {Marconi},
  {Gnerucci}, \& {Magrini}}]{cresci10}
{Cresci}, G., {Mannucci}, F., {Maiolino}, R., {et~al.} 2010, \nat, 467, 811,
  \dodoi{10.1038/nature09451}

\bibitem[{{Curti} {et~al.}(2017){Curti}, {Cresci}, {Mannucci}, {Marconi},
  {Maiolino}, \& {Esposito}}]{curti16}
{Curti}, M., {Cresci}, G., {Mannucci}, F., {et~al.} 2017, \mnras, 465, 1384,
  \dodoi{10.1093/mnras/stw2766}

\bibitem[{{Curti} {et~al.}(2020){Curti}, {Mannucci}, {Cresci}, \&
  {Maiolino}}]{curti20}
{Curti}, M., {Mannucci}, F., {Cresci}, G., \& {Maiolino}, R. 2020, \mnras, 491,
  944, \dodoi{10.1093/mnras/stz2910}

\bibitem[{{da Cunha} {et~al.}(2008){da Cunha}, {Charlot}, \&
  {Elbaz}}]{dacunha08}
{da Cunha}, E., {Charlot}, S., \& {Elbaz}, D. 2008, \mnras, 388, 1595,
  \dodoi{10.1111/j.1365-2966.2008.13535.x}

\bibitem[{{da Cunha} {et~al.}(2015){da Cunha}, {Walter}, {Smail}, {Swinbank},
  {Simpson}, {Decarli}, {Hodge}, {Weiss}, {van der Werf}, {Bertoldi},
  {Chapman}, {Cox}, {Danielson}, {Dannerbauer}, {Greve}, {Ivison}, {Karim}, \&
  {Thomson}}]{dacunha15}
{da Cunha}, E., {Walter}, F., {Smail}, I.~R., {et~al.} 2015, \apj, 806, 110,
  \dodoi{10.1088/0004-637X/806/1/110}

\bibitem[{{Daddi} {et~al.}(2010){Daddi}, {Elbaz}, {Walter}, {Bournaud},
  {Salmi}, {Carilli}, {Dannerbauer}, {Dickinson}, {Monaco}, \&
  {Riechers}}]{daddi10}
{Daddi}, E., {Elbaz}, D., {Walter}, F., {et~al.} 2010, \apjl, 714, L118,
  \dodoi{10.1088/2041-8205/714/1/L118}

\bibitem[{{Daddi} {et~al.}(2015){Daddi}, {Dannerbauer}, {Liu}, {Aravena},
  {Bournaud}, {Walter}, {Riechers}, {Magdis}, {Sargent}, {B{\'e}thermin},
  {Carilli}, {Cibinel}, {Dickinson}, {Elbaz}, {Gao}, {Gobat}, {Hodge}, \&
  {Krips}}]{daddi15}
{Daddi}, E., {Dannerbauer}, H., {Liu}, D., {et~al.} 2015, \aap, 577, A46,
  \dodoi{10.1051/0004-6361/201425043}

\bibitem[{{Dav{\'e}} {et~al.}(2012){Dav{\'e}}, {Finlator}, \&
  {Oppenheimer}}]{dave12}
{Dav{\'e}}, R., {Finlator}, K., \& {Oppenheimer}, B.~D. 2012, \mnras, 421, 98,
  \dodoi{10.1111/j.1365-2966.2011.20148.x}

\bibitem[{{Dav{\'e}} {et~al.}(2011){Dav{\'e}}, {Oppenheimer}, \&
  {Finlator}}]{dave11_1}
{Dav{\'e}}, R., {Oppenheimer}, B.~D., \& {Finlator}, K. 2011, \mnras, 415, 11,
  \dodoi{10.1111/j.1365-2966.2011.18680.x}

\bibitem[{{Davies} {et~al.}(2019){Davies}, {F{\"o}rster Schreiber},
  {{\"U}bler}, {Genzel}, {Lutz}, {Renzini}, {Tacchella}, {Tacconi}, {Belli},
  {Burkert}, {Carollo}, {Davies}, {Herrera-Camus}, {Lilly}, {Mancini}, {Naab},
  {Nelson}, {Price}, {Shimizu}, {Sternberg}, {Wisnioski}, \&
  {Wuyts}}]{davies19}
{Davies}, R.~L., {F{\"o}rster Schreiber}, N.~M., {{\"U}bler}, H., {et~al.}
  2019, \apj, 873, 122, \dodoi{10.3847/1538-4357/ab06f1}

\bibitem[{{Dayal} {et~al.}(2013){Dayal}, {Ferrara}, \& {Dunlop}}]{Dayal13}
{Dayal}, P., {Ferrara}, A., \& {Dunlop}, J.~S. 2013, \mnras, 430, 2891,
  \dodoi{10.1093/mnras/stt083}

\bibitem[{{Draine} \& {Li}(2007)}]{draineli07}
{Draine}, B.~T., \& {Li}, A. 2007, \apj, 657, 810, \dodoi{10.1086/511055}

\bibitem[{{Dudzevi{\v{c}}i{\={u}}t{\.{e}}}
  {et~al.}(2020){Dudzevi{\v{c}}i{\={u}}t{\.{e}}}, {Smail}, {Swinbank}, {Stach},
  {Almaini}, {da Cunha}, {An}, {Arumugam}, {Birkin}, {Blain}, {Chapman},
  {Chen}, {Conselice}, {Coppin}, {Dunlop}, {Farrah}, {Geach}, {Gullberg},
  {Hartley}, {Hodge}, {Ivison}, {Maltby}, {Scott}, {Simpson}, {Simpson},
  {Thomson}, {Walter}, {Wardlow}, {Weiss}, \& {van der Werf}}]{dudzeviciute20}
{Dudzevi{\v{c}}i{\={u}}t{\.{e}}}, U., {Smail}, I., {Swinbank}, A.~M., {et~al.}
  2020, \mnras, 494, 3828, \dodoi{10.1093/mnras/staa769}

\bibitem[{{Elbaz} {et~al.}(2018){Elbaz}, {Leiton}, {Nagar}, {Okumura},
  {Franco}, {Schreiber}, {Pannella}, {Wang}, {Dickinson}, {D{\'\i}az-Santos},
  {Ciesla}, {Daddi}, {Bournaud}, {Magdis}, {Zhou}, \& {Rujopakarn}}]{elbaz18}
{Elbaz}, D., {Leiton}, R., {Nagar}, N., {et~al.} 2018, \aap, 616, A110,
  \dodoi{10.1051/0004-6361/201732370}

\bibitem[{{Elvis} {et~al.}(2009){Elvis}, {Civano}, {Vignali}, {Puccetti},
  {Fiore}, {Cappelluti}, {Aldcroft}, {Fruscione}, {Zamorani}, {Comastri},
  {Brusa}, {Gilli}, {Miyaji}, {Damiani}, {Koekemoer}, {Finoguenov}, {Brunner},
  {Urry}, {Silverman}, {Mainieri}, {Hasinger}, {Griffiths}, {Carollo}, {Hao},
  {Guzzo}, {Blain}, {Calzetti}, {Carilli}, {Capak}, {Ettori}, {Fabbiano},
  {Impey}, {Lilly}, {Mobasher}, {Rich}, {Salvato}, {Sand ers}, {Schinnerer},
  {Scoville}, {Shopbell}, {Taylor}, {Taniguchi}, \& {Volonteri}}]{elvis09}
{Elvis}, M., {Civano}, F., {Vignali}, C., {et~al.} 2009, \apjs, 184, 158,
  \dodoi{10.1088/0067-0049/184/1/158}

\bibitem[{{Erb}(2008)}]{erb08_apj674}
{Erb}, D.~K. 2008, \apj, 674, 151, \dodoi{10.1086/524727}

\bibitem[{{Faucher-Gigu{\`e}re}(2018)}]{faucher-giguere18}
{Faucher-Gigu{\`e}re}, C.-A. 2018, \mnras, 473, 3717,
  \dodoi{10.1093/mnras/stx2595}

\bibitem[{{Faucher-Gigu{\`e}re} {et~al.}(2011){Faucher-Gigu{\`e}re},
  {Kere{\v{s}}}, \& {Ma}}]{faucher-giguere11}
{Faucher-Gigu{\`e}re}, C.-A., {Kere{\v{s}}}, D., \& {Ma}, C.-P. 2011, \mnras,
  417, 2982, \dodoi{10.1111/j.1365-2966.2011.19457.x}

\bibitem[{{Feldmann} {et~al.}(2006){Feldmann}, {Carollo}, {Porciani}, {Lilly},
  {Capak}, {Taniguchi}, {Le F{\`e}vre}, {Renzini}, {Scoville}, {Ajiki},
  {Aussel}, {Contini}, {McCracken}, {Mobasher}, {Murayama}, {Sanders},
  {Sasaki}, {Scarlata}, {Scodeggio}, {Shioya}, {Silverman}, {Takahashi},
  {Thompson}, \& {Zamorani}}]{feldmann06}
{Feldmann}, R., {Carollo}, C.~M., {Porciani}, C., {et~al.} 2006, \mnras, 372,
  565, \dodoi{10.1111/j.1365-2966.2006.10930.x}

\bibitem[{{Finlator} \& {Dav{\'e}}(2008)}]{finlator08}
{Finlator}, K., \& {Dav{\'e}}, R. 2008, \mnras, 385, 2181,
  \dodoi{10.1111/j.1365-2966.2008.12991.x}

\bibitem[{{Freundlich} {et~al.}(2019){Freundlich}, {Combes}, {Tacconi},
  {Genzel}, {Garcia-Burillo}, {Neri}, {Contini}, {Bolatto}, {Lilly},
  {Salom{\'e}}, {Bicalho}, {Boissier}, {Boone}, {Bouch{\'e}}, {Bournaud},
  {Burkert}, {Carollo}, {Cooper}, {Cox}, {Feruglio}, {F{\"o}rster Schreiber},
  {Juneau}, {Lippa}, {Lutz}, {Naab}, {Renzini}, {Saintonge}, {Sternberg},
  {Walter}, {Weiner}, {Wei{\ss}}, \& {Wuyts}}]{freundlich19}
{Freundlich}, J., {Combes}, F., {Tacconi}, L.~J., {et~al.} 2019, \aap, 622,
  A105, \dodoi{10.1051/0004-6361/201732223}

\bibitem[{{Geach} {et~al.}(2011){Geach}, {Smail}, {Moran}, {MacArthur},
  {Lagos}, \& {Edge}}]{geach11}
{Geach}, J.~E., {Smail}, I., {Moran}, S.~M., {et~al.} 2011, \apjl, 730, L19,
  \dodoi{10.1088/2041-8205/730/2/L19}

\bibitem[{{Genzel} {et~al.}(2010){Genzel}, {Tacconi}, {Gracia-Carpio},
  {Sternberg}, {Cooper}, {Shapiro}, {Bolatto}, {Bouch{\'e}}, {Bournaud},
  {Burkert}, {Combes}, {Comerford}, {Cox}, {Davis}, {Schreiber},
  {Garcia-Burillo}, {Lutz}, {Naab}, {Neri}, {Omont}, {Shapley}, \&
  {Weiner}}]{genzel10}
{Genzel}, R., {Tacconi}, L.~J., {Gracia-Carpio}, J., {et~al.} 2010, \mnras,
  407, 2091, \dodoi{10.1111/j.1365-2966.2010.16969.x}

\bibitem[{{Genzel} {et~al.}(2011){Genzel}, {Newman}, {Jones}, {F{\"o}rster
  Schreiber}, {Shapiro}, {Genel}, {Lilly}, {Renzini}, {Tacconi}, {Bouch{\'e}},
  {Burkert}, {Cresci}, {Buschkamp}, {Carollo}, {Ceverino}, {Davies}, {Dekel},
  {Eisenhauer}, {Hicks}, {Kurk}, {Lutz}, {Mancini}, {Naab}, {Peng},
  {Sternberg}, {Vergani}, \& {Zamorani}}]{genzel11}
{Genzel}, R., {Newman}, S., {Jones}, T., {et~al.} 2011, \apj, 733, 101,
  \dodoi{10.1088/0004-637X/733/2/101}

\bibitem[{{Genzel} {et~al.}(2012){Genzel}, {Tacconi}, {Combes}, {Bolatto},
  {Neri}, {Sternberg}, {Cooper}, {Bouch{\'e}}, {Bournaud}, {Burkert},
  {Comerford}, {Cox}, {Davis}, {F{\"o}rster Schreiber}, {Garcia-Burillo},
  {Gracia-Carpio}, {Lutz}, {Naab}, {Newman}, {Saintonge}, {Shapiro}, {Shapley},
  \& {Weiner}}]{genzel12}
{Genzel}, R., {Tacconi}, L.~J., {Combes}, F., {et~al.} 2012, \apj, 746, 69,
  \dodoi{10.1088/0004-637X/746/1/69}

\bibitem[{{Groves} {et~al.}(2015){Groves}, {Schinnerer}, {Leroy}, {Galametz},
  {Walter}, {Bolatto}, {Hunt}, {Dale}, {Calzetti}, {Croxall}, \&
  {Kennicutt}}]{groves15}
{Groves}, B.~A., {Schinnerer}, E., {Leroy}, A., {et~al.} 2015, \apj, 799, 96,
  \dodoi{10.1088/0004-637X/799/1/96}

\bibitem[{{Guo} {et~al.}(2016){Guo}, {Rafelski}, {Faber}, {Koo}, {Krumholz},
  {Trump}, {Willner}, {Amor{\'\i}n}, {Barro}, {Bell}, {Gardner}, {Gawiser},
  {Hathi}, {Koekemoer}, {Pacifici}, {P{\'e}rez-Gonz{\'a}lez}, {Ravindranath},
  {Reddy}, {Teplitz}, \& {Yesuf}}]{yicheng16}
{Guo}, Y., {Rafelski}, M., {Faber}, S.~M., {et~al.} 2016, \apj, 833, 37,
  \dodoi{10.3847/1538-4357/833/1/37}

\bibitem[{{Hayward} \& {Hopkins}(2017)}]{hayword17}
{Hayward}, C.~C., \& {Hopkins}, P.~F. 2017, \mnras, 465, 1682,
  \dodoi{10.1093/mnras/stw2888}
  
\bibitem[{{Herrera-Camus} {et~al.}(2018){Herrera-Camus}, {Sturm},
  {Graci{\'a}-Carpio}, {Lutz}, {Contursi}, {Veilleux}, {Fischer},
  {Gonz{\'a}lez-Alfonso}, {Poglitsch}, {Tacconi}, {Genzel}, {Maiolino},
  {Sternberg}, {Davies}, \& {Verma}}]{herrera-camus18}
{Herrera-Camus}, R., {Sturm}, E., {Graci{\'a}-Carpio}, J., {et~al.} 2018, \apj,
  861, 94, \dodoi{10.3847/1538-4357/aac0f6}

\bibitem[{{Hunt} {et~al.}(2015){Hunt}, {Garc{\'\i}a-Burillo}, {Casasola},
  {Caselli}, {Combes}, {Henkel}, {Lundgren}, {Maiolino}, {Menten}, {Testi}, \&
  {Weiss}}]{hunt15}
{Hunt}, L.~K., {Garc{\'\i}a-Burillo}, S., {Casasola}, V., {et~al.} 2015, \aap,
  583, A114, \dodoi{10.1051/0004-6361/201526553}

\bibitem[{{Hunt} {et~al.}(2019){Hunt}, {De Looze}, {Boquien}, {Nikutta},
  {Rossi}, {Bianchi}, {Dale}, {Granato}, {Kennicutt}, {Silva}, {Ciesla},
  {Rela{\~n}o}, {Viaene}, {Brandl}, {Calzetti}, {Croxall}, {Draine},
  {Galametz}, {Gordon}, {Groves}, {Helou}, {Herrera-Camus}, {Hinz}, {Koda},
  {Salim}, {Sandstrom}, {Smith}, {Wilson}, \& {Zibetti}}]{hunt19}
{Hunt}, L.~K., {De Looze}, I., {Boquien}, M., {et~al.} 2019, \aap, 621, A51,
  \dodoi{10.1051/0004-6361/201834212}

\bibitem[{{Ilbert} {et~al.}(2013){Ilbert}, {McCracken}, {Le F{\`e}vre},
  {Capak}, {Dunlop}, {Karim}, {Renzini}, {Caputi}, {Boissier}, {Arnouts},
  {Aussel}, {Comparat}, {Guo}, {Hudelot}, {Kartaltepe}, {Kneib}, {Krogager},
  {Le Floc'h}, {Lilly}, {Mellier}, {Milvang-Jensen}, {Moutard}, {Onodera},
  {Richard}, {Salvato}, {Sanders}, {Scoville}, {Silverman}, {Taniguchi},
  {Tasca}, {Thomas}, {Toft}, {Tresse}, {Vergani}, {Wolk}, \& {Zirm}}]{ilbert13}
{Ilbert}, O., {McCracken}, H.~J., {Le F{\`e}vre}, O., {et~al.} 2013, \aap, 556,
  A55, \dodoi{10.1051/0004-6361/201321100}

\bibitem[{{Kalfountzou} {et~al.}(2014){Kalfountzou}, {Civano}, {Elvis},
  {Trichas}, \& {Green}}]{kalfountzou14}
{Kalfountzou}, E., {Civano}, F., {Elvis}, M., {Trichas}, M., \& {Green}, P.
  2014, \mnras, 445, 1430, \dodoi{10.1093/mnras/stu1745}

\bibitem[{{Kashino} \& {Inoue}(2019)}]{kashino19}
{Kashino}, D., \& {Inoue}, A.~K. 2019, \mnras, 486, 1053,
  \dodoi{10.1093/mnras/stz881}

\bibitem[{{Kennicutt}(1998)}]{kennicutt98}
{Kennicutt}, Robert~C., J. 1998, \apj, 498, 541, \dodoi{10.1086/305588}

\bibitem[{{Kewley} \& {Dopita}(2002)}]{kewleydopita02}
{Kewley}, L.~J., \& {Dopita}, M.~A. 2002, \apjs, 142, 35,
  \dodoi{10.1086/341326}

\bibitem[{{Khostovan} {et~al.}(2015){Khostovan}, {Sobral}, {Mobasher}, {Best},
  {Smail}, {Stott}, {Hemmati}, \& {Nayyeri}}]{khostovan15}
{Khostovan}, A.~A., {Sobral}, D., {Mobasher}, B., {et~al.} 2015, \mnras, 452,
  3948, \dodoi{10.1093/mnras/stv1474}

\bibitem[{{Kobulnicky} \& {Kewley}(2004)}]{KK04}
{Kobulnicky}, H.~A., \& {Kewley}, L.~J. 2004, \apj, 617, 240,
  \dodoi{10.1086/425299}

\bibitem[{{Kriek} {et~al.}(2009){Kriek}, {van Dokkum}, {Labb{\'e}}, {Franx},
  {Illingworth}, {Marchesini}, \& {Quadri}}]{kriek09b}
{Kriek}, M., {van Dokkum}, P.~G., {Labb{\'e}}, I., {et~al.} 2009, \apj, 700,
  221, \dodoi{10.1088/0004-637X/700/1/221}

\bibitem[{{Lagos} {et~al.}(2016){Lagos}, {Theuns}, {Schaye}, {Furlong},
  {Bower}, {Schaller}, {Crain}, {Trayford}, \& {Matthee}}]{lagos16}
{Lagos}, C. d.~P., {Theuns}, T., {Schaye}, J., {et~al.} 2016, \mnras, 459,
  2632, \dodoi{10.1093/mnras/stw717}

\bibitem[{{Laigle} {et~al.}(2016){Laigle}, {McCracken}, {Ilbert}, {Hsieh},
  {Davidzon}, {Capak}, {Hasinger}, {Silverman}, {Pichon}, {Coupon}, {Aussel},
  {Le Borgne}, {Caputi}, {Cassata}, {Chang}, {Civano}, {Dunlop}, {Fynbo},
  {Kartaltepe}, {Koekemoer}, {Le F{\`e}vre}, {Le Floc'h}, {Leauthaud}, {Lilly},
  {Lin}, {Marchesi}, {Milvang-Jensen}, {Salvato}, {Sanders}, {Scoville},
  {Smolcic}, {Stockmann}, {Taniguchi}, {Tasca}, {Toft}, {Vaccari}, \&
  {Zabl}}]{cosmos2015}
{Laigle}, C., {McCracken}, H.~J., {Ilbert}, O., {et~al.} 2016, \apjs, 224, 24,
  \dodoi{10.3847/0067-0049/224/2/24}

\bibitem[{{Leroy} {et~al.}(2011){Leroy}, {Bolatto}, {Gordon}, {Sand strom},
  {Gratier}, {Rosolowsky}, {Engelbracht}, {Mizuno}, {Corbelli}, {Fukui}, \&
  {Kawamura}}]{leroy11}
{Leroy}, A.~K., {Bolatto}, A., {Gordon}, K., {et~al.} 2011, \apj, 737, 12,
  \dodoi{10.1088/0004-637X/737/1/12}

\bibitem[{{Lilly} {et~al.}(2013){Lilly}, {Carollo}, {Pipino}, {Renzini}, \&
  {Peng}}]{lilly13}
{Lilly}, S.~J., {Carollo}, C.~M., {Pipino}, A., {Renzini}, A., \& {Peng}, Y.
  2013, \apj, 772, 119, \dodoi{10.1088/0004-637X/772/2/119}

\bibitem[{{Lilly} {et~al.}(2007){Lilly}, {Le F{\`e}vre}, {Renzini}, {Zamorani},
  {Scodeggio}, {Contini}, {Carollo}, {Hasinger}, {Kneib}, {Iovino}, {Le Brun},
  {Maier}, {Mainieri}, {Mignoli}, {Silverman}, {Tasca}, {Bolzonella},
  {Bongiorno}, {Bottini}, {Capak}, {Caputi}, {Cimatti}, {Cucciati}, {Daddi},
  {Feldmann}, {Franzetti}, {Garilli}, {Guzzo}, {Ilbert}, {Kampczyk}, {Kovac},
  {Lamareille}, {Leauthaud}, {Le Borgne}, {McCracken}, {Marinoni}, {Pello},
  {Ricciardelli}, {Scarlata}, {Vergani}, {Sanders}, {Schinnerer}, {Scoville},
  {Taniguchi}, {Arnouts}, {Aussel}, {Bardelli}, {Brusa}, {Cappi}, {Ciliegi},
  {Finoguenov}, {Foucaud}, {Franceschini}, {Halliday}, {Impey}, {Knobel},
  {Koekemoer}, {Kurk}, {Maccagni}, {Maddox}, {Marano}, {Marconi}, {Meneux},
  {Mobasher}, {Moreau}, {Peacock}, {Porciani}, {Pozzetti}, {Scaramella},
  {Schiminovich}, {Shopbell}, {Smail}, {Thompson}, {Tresse}, {Vettolani},
  {Zanichelli}, \& {Zucca}}]{lilly07}
{Lilly}, S.~J., {Le F{\`e}vre}, O., {Renzini}, A., {et~al.} 2007, \apjs, 172,
  70, \dodoi{10.1086/516589}

\bibitem[{{Liu} {et~al.}(2019){Liu}, {Schinnerer}, {Groves}, {Magnelli},
  {Lang}, {Leslie}, {Jim{\'e}nez-Andrade}, {Riechers}, {Popping}, {Magdis},
  {Daddi}, {Sargent}, {Gao}, {Fudamoto}, {Oesch}, \& {Bertoldi}}]{liu19_II}
{Liu}, D., {Schinnerer}, E., {Groves}, B., {et~al.} 2019, \apj, 887, 235,
  \dodoi{10.3847/1538-4357/ab578d}

\bibitem[{{Madau} \& {Dickinson}(2014)}]{madau14}
{Madau}, P., \& {Dickinson}, M. 2014, \araa, 52, 415,
  \dodoi{10.1146/annurev-astro-081811-125615}

\bibitem[{{Magdis} {et~al.}(2012){Magdis}, {Daddi}, {B{\'e}thermin}, {Sargent},
  {Elbaz}, {Pannella}, {Dickinson}, {Dannerbauer}, {da Cunha}, {Walter},
  {Rigopoulou}, {Charmandaris}, {Hwang}, \& {Kartaltepe}}]{magdis12}
{Magdis}, G.~E., {Daddi}, E., {B{\'e}thermin}, M., {et~al.} 2012, \apj, 760, 6,
  \dodoi{10.1088/0004-637X/760/1/6}

\bibitem[{{Magdis} {et~al.}(2017){Magdis}, {Rigopoulou}, {Daddi}, {Bethermin},
  {Feruglio}, {Sargent}, {Dannerbauer}, {Dickinson}, {Elbaz}, {Gomez Guijarro},
  {Huang}, {Toft}, \& {Valentino}}]{magdis17}
{Magdis}, G.~E., {Rigopoulou}, D., {Daddi}, E., {et~al.} 2017, \aap, 603, A93,
  \dodoi{10.1051/0004-6361/201731037}

\bibitem[{{Mannucci} {et~al.}(2010){Mannucci}, {Cresci}, {Maiolino}, {Marconi},
  \& {Gnerucci}}]{mannucci10}
{Mannucci}, F., {Cresci}, G., {Maiolino}, R., {Marconi}, A., \& {Gnerucci}, A.
  2010, \mnras, 408, 2115, \dodoi{10.1111/j.1365-2966.2010.17291.x}

\bibitem[{{McCracken} {et~al.}(2012){McCracken}, {Milvang-Jensen}, {Dunlop},
  {Franx}, {Fynbo}, {Le F{\`e}vre}, {Holt}, {Caputi}, {Goranova}, {Buitrago},
  {Emerson}, {Freudling}, {Hudelot}, {L{\'o}pez-Sanjuan}, {Magnard}, {Mellier},
  {M{\o}ller}, {Nilsson}, {Sutherland}, {Tasca}, \& {Zabl}}]{mccracken12}
{McCracken}, H.~J., {Milvang-Jensen}, B., {Dunlop}, J., {et~al.} 2012, \aap,
  544, A156, \dodoi{10.1051/0004-6361/201219507}

\bibitem[{{McLean} {et~al.}(2010){McLean}, {Steidel}, {Epps}, {Matthews},
  {Adkins}, {Konidaris}, {Weber}, {Aliado}, {Brims}, {Canfield}, {Cromer},
  {Fucik}, {Kulas}, {Mace}, {Magnone}, {Rodriguez}, {Wang}, \&
  {Weiss}}]{mclean10}
{McLean}, I.~S., {Steidel}, C.~C., {Epps}, H., {et~al.} 2010, in \procspie,
  Vol. 7735, Ground-based and Airborne Instrumentation for Astronomy III,
  77351E--77351E--12, \dodoi{10.1117/12.856715}

\bibitem[{{McLean} {et~al.}(2012){McLean}, {Steidel}, {Epps}, {Konidaris},
  {Matthews}, {Adkins}, {Aliado}, {Brims}, {Canfield}, {Cromer}, {Fucik},
  {Kulas}, {Mace}, {Magnone}, {Rodriguez}, {Rudie}, {Trainor}, {Wang}, {Weber},
  \& {Weiss}}]{mclean12}
{McLean}, I.~S., {Steidel}, C.~C., {Epps}, H.~W., {et~al.} 2012, in \procspie,
  Vol. 8446, Ground-based and Airborne Instrumentation for Astronomy IV,
  84460J, \dodoi{10.1117/12.924794}

\bibitem[{{McMullin} {et~al.}(2007){McMullin}, {Waters}, {Schiebel}, {Young},
  \& {Golap}}]{CASA}
{McMullin}, J.~P., {Waters}, B., {Schiebel}, D., {Young}, W., \& {Golap}, K.
  2007, Astronomical Society of the Pacific Conference Series, Vol. 376, {CASA
  Architecture and Applications}, ed. R.~A. {Shaw}, F.~{Hill}, \& D.~J. {Bell},
  127

\bibitem[{{Mitra} {et~al.}(2015){Mitra}, {Dav{\'e}}, \& {Finlator}}]{mitra15}
{Mitra}, S., {Dav{\'e}}, R., \& {Finlator}, K. 2015, \mnras, 452, 1184,
  \dodoi{10.1093/mnras/stv1387}

\bibitem[{{Muratov} {et~al.}(2015){Muratov}, {Kere{\v{s}}},
  {Faucher-Gigu{\`e}re}, {Hopkins}, {Quataert}, \& {Murray}}]{muratov15}
{Muratov}, A.~L., {Kere{\v{s}}}, D., {Faucher-Gigu{\`e}re}, C.-A., {et~al.}
  2015, \mnras, 454, 2691, \dodoi{10.1093/mnras/stv2126}

%\bibitem[{{Narayanan} \& {Krumholz}(2014)}]{narayanan14}
%{Narayanan}, D., \& {Krumholz}, M.~R. 2014, \mnras, 442, 1411,
%  \dodoi{10.1093/mnras/stu834}

\bibitem[{{Onodera} {et~al.}(2016){Onodera}, {Carollo}, {Lilly}, {Renzini},
  {Arimoto}, {Capak}, {Daddi}, {Scoville}, {Tacchella}, {Tatehora}, \&
  {Zamorani}}]{onodera16}
{Onodera}, M., {Carollo}, C.~M., {Lilly}, S., {et~al.} 2016, \apj, 822, 42,
  \dodoi{10.3847/0004-637X/822/1/42}

\bibitem[{{Peng} \& {Maiolino}(2014)}]{pengmaiolino14}
{Peng}, Y.-j., \& {Maiolino}, R. 2014, \mnras, 443, 3643,
  \dodoi{10.1093/mnras/stu1288}

\bibitem[{{Pettini} \& {Pagel}(2004)}]{PP04}
{Pettini}, M., \& {Pagel}, B.~E.~J. 2004, \mnras, 348, L59,
  \dodoi{10.1111/j.1365-2966.2004.07591.x}

\bibitem[{{Popping} {et~al.}(2015){Popping}, {Caputi}, {Trager}, {Somerville},
  {Dekel}, {Kassin}, {Kocevski}, {Koekemoer}, {Faber}, {Ferguson}, {Galametz},
  {Grogin}, {Guo}, {Lu}, {Wel}, \& {Weiner}}]{popping15}
{Popping}, G., {Caputi}, K.~I., {Trager}, S.~C., {et~al.} 2015, \mnras, 454,
  2258, \dodoi{10.1093/mnras/stv2136}

\bibitem[{{Price-Whelan} {et~al.}(2018){Price-Whelan}, {Sip{\H{o}}cz},
  {G{\"u}nther}, {Lim}, {Crawford}, {Conseil}, {Shupe}, {Craig}, {Dencheva},
  {Ginsburg}, {VanderPlas}, {Bradley}, {P{\'e}rez-Su{\'a}rez}, {de Val-Borro},
  {Paper Contributors}, {Aldcroft}, {Cruz}, {Robitaille}, {Tollerud},
  {Coordination Committee}, {Ardelean}, {Babej}, {Bach}, {Bachetti}, {Bakanov},
  {Bamford}, {Barentsen}, {Barmby}, {Baumbach}, {Berry}, {Biscani}, {Boquien},
  {Bostroem}, {Bouma}, {Brammer}, {Bray}, {Breytenbach}, {Buddelmeijer},
  {Burke}, {Calderone}, {Cano Rodr{\'\i}guez}, {Cara}, {Cardoso}, {Cheedella},
  {Copin}, {Corrales}, {Crichton}, {D{\textquoteright}Avella}, {Deil},
  {Depagne}, {Dietrich}, {Donath}, {Droettboom}, {Earl}, {Erben}, {Fabbro},
  {Ferreira}, {Finethy}, {Fox}, {Garrison}, {Gibbons}, {Goldstein}, {Gommers},
  {Greco}, {Greenfield}, {Groener}, {Grollier}, {Hagen}, {Hirst}, {Homeier},
  {Horton}, {Hosseinzadeh}, {Hu}, {Hunkeler}, {Ivezi{\'c}}, {Jain}, {Jenness},
  {Kanarek}, {Kendrew}, {Kern}, {Kerzendorf}, {Khvalko}, {King}, {Kirkby},
  {Kulkarni}, {Kumar}, {Lee}, {Lenz}, {Littlefair}, {Ma}, {Macleod},
  {Mastropietro}, {McCully}, {Montagnac}, {Morris}, {Mueller}, {Mumford},
  {Muna}, {Murphy}, {Nelson}, {Nguyen}, {Ninan}, {N{\"o}the}, {Ogaz}, {Oh},
  {Parejko}, {Parley}, {Pascual}, {Patil}, {Patil}, {Plunkett}, {Prochaska},
  {Rastogi}, {Reddy Janga}, {Sabater}, {Sakurikar}, {Seifert}, {Sherbert},
  {Sherwood-Taylor}, {Shih}, {Sick}, {Silbiger}, {Singanamalla}, {Singer},
  {Sladen}, {Sooley}, {Sornarajah}, {Streicher}, {Teuben}, {Thomas},
  {Tremblay}, {Turner}, {Terr{\'o}n}, {van Kerkwijk}, {de la Vega}, {Watkins},
  {Weaver}, {Whitmore}, {Woillez}, {Zabalza}, \& {Contributors}}]{astropy:2018}
{Price-Whelan}, A.~M., {Sip{\H{o}}cz}, B.~M., {G{\"u}nther}, H.~M., {et~al.}
  2018, \aj, 156, 123, \dodoi{10.3847/1538-3881/aabc4f}

\bibitem[{{R{\'e}my-Ruyer} {et~al.}(2014){R{\'e}my-Ruyer}, {Madden},
  {Galliano}, {Galametz}, {Takeuchi}, {Asano}, {Zhukovska}, {Lebouteiller},
  {Cormier}, {Jones}, {Bocchio}, {Baes}, {Bendo}, {Boquien}, {Boselli},
  {DeLooze}, {Doublier-Pritchard}, {Hughes}, {Karczewski}, \&
  {Spinoglio}}]{remy-ruyer14}
{R{\'e}my-Ruyer}, A., {Madden}, S.~C., {Galliano}, F., {et~al.} 2014, \aap,
  563, A31, \dodoi{10.1051/0004-6361/201322803}

\bibitem[{{Riechers} {et~al.}(2019){Riechers}, {Pavesi}, {Sharon}, {Hodge},
  {Decarli}, {Walter}, {Carilli}, {Aravena}, {da Cunha}, {Daddi}, {Dickinson},
  {Smail}, {Capak}, {Ivison}, {Sargent}, {Scoville}, \& {Wagg}}]{riechers19}
{Riechers}, D.~A., {Pavesi}, R., {Sharon}, C.~E., {et~al.} 2019, \apj, 872, 7,
  \dodoi{10.3847/1538-4357/aafc27}

\bibitem[{{Riechers} {et~al.}(2020){Riechers}, {Boogaard}, {Decarli},
  {Gonz{\'a}lez-L{\'o}pez}, {Smail}, {Walter}, {Aravena}, {Carilli}, {Cortes},
  {Cox}, {D{\'\i}az-Santos}, {Hodge}, {Inami}, {Ivison}, {Kaasinen}, {Wagg},
  {Wei{\ss}}, \& {van der Werf}}]{riechers20}
{Riechers}, D.~A., {Boogaard}, L.~A., {Decarli}, R., {et~al.} 2020, \apjl, 896,
  L21, \dodoi{10.3847/2041-8213/ab9595}

\bibitem[{{Rupke} {et~al.}(2008){Rupke}, {Veilleux}, \& {Baker}}]{rupke08}
{Rupke}, D. S.~N., {Veilleux}, S., \& {Baker}, A.~J. 2008, \apj, 674, 172,
  \dodoi{10.1086/522363}

\bibitem[{{Saintonge} {et~al.}(2012){Saintonge}, {Tacconi}, {Fabello}, {Wang},
  {Catinella}, {Genzel}, {Graci{\'a}-Carpio}, {Kramer}, {Moran}, {Heckman},
  {Schiminovich}, {Schuster}, \& {Wuyts}}]{saintonge12}
{Saintonge}, A., {Tacconi}, L.~J., {Fabello}, S., {et~al.} 2012, \apj, 758, 73,
  \dodoi{10.1088/0004-637X/758/2/73}

\bibitem[{{Saintonge} {et~al.}(2013){Saintonge}, {Lutz}, {Genzel}, {Magnelli},
  {Nordon}, {Tacconi}, {Baker}, {Bandara}, {Berta}, {F{\"o}rster Schreiber},
  {Poglitsch}, {Sturm}, {Wuyts}, \& {Wuyts}}]{saintonge13}
{Saintonge}, A., {Lutz}, D., {Genzel}, R., {et~al.} 2013, \apj, 778, 2,
  \dodoi{10.1088/0004-637X/778/1/2}

\bibitem[{{Saintonge} {et~al.}(2017){Saintonge}, {Catinella}, {Tacconi},
  {Kauffmann}, {Genzel}, {Cortese}, {Dav{\'e}}, {Fletcher},
  {Graci{\'a}-Carpio}, {Kramer}, {Heckman}, {Janowiecki}, {Lutz}, {Rosario},
  {Schiminovich}, {Schuster}, {Wang}, {Wuyts}, {Borthakur}, {Lamperti}, \&
  {Roberts-Borsani}}]{saintonge17}
{Saintonge}, A., {Catinella}, B., {Tacconi}, L.~J., {et~al.} 2017, \apjs, 233,
  22, \dodoi{10.3847/1538-4365/aa97e0}
  
\bibitem[{{Sanders} {et~al.}(2020){Sanders}, {Shapley}, {Jones}, {Reddy},
  {Kriek}, {Siana}, {Coil}, {Mobasher}, {Shivaei}, {Dav{\'e}}, {Azadi},
  {Price}, {Leung}, {Freeman}, {Fetherolf}, {de Groot}, {Zick}, \&
  {Barro}}]{sanders20}
{Sanders}, R.~L., {Shapley}, A.~E., {Jones}, T., {et~al.} 2020, arXiv e-prints,
  arXiv:2009.07292.
\newblock \doarXiv{2009.07292}

\bibitem[{{Santini} {et~al.}(2010){Santini}, {Maiolino}, {Magnelli}, {Silva},
  {Grazian}, {Altieri}, {Andreani}, {Aussel}, {Berta}, {Bongiovanni},
  {Brisbin}, {Calura}, {Cava}, {Cepa}, {Cimatti}, {Daddi}, {Dannerbauer},
  {Dominguez-Sanchez}, {Elbaz}, {Fontana}, {F{\"o}rster Schreiber}, {Genzel},
  {Granato}, {Gruppioni}, {Lutz}, {Magdis}, {Magliocchetti}, {Matteucci},
  {Nordon}, {P{\'e}rez Garcia}, {Poglitsch}, {Popesso}, {Pozzi}, {Riguccini},
  {Rodighiero}, {Saintonge}, {Sanchez-Portal}, {Shao}, {Sturm}, {Tacconi}, \&
  {Valtchanov}}]{santini10}
{Santini}, P., {Maiolino}, R., {Magnelli}, B., {et~al.} 2010, \aap, 518, L154,
  \dodoi{10.1051/0004-6361/201014748}


\bibitem[{{Santini} {et~al.}(2014){Santini}, {Maiolino}, {Magnelli}, {Lutz},
  {Lamastra}, {Li Causi}, {Eales}, {Andreani}, {Berta}, {Buat}, {Cooray},
  {Cresci}, {Daddi}, {Farrah}, {Fontana}, {Franceschini}, {Genzel}, {Granato},
  {Grazian}, {Le Floc'h}, {Magdis}, {Magliocchetti}, {Mannucci}, {Menci},
  {Nordon}, {Oliver}, {Popesso}, {Pozzi}, {Riguccini}, {Rodighiero}, {Rosario},
  {Salvato}, {Scott}, {Silva}, {Tacconi}, {Viero}, {Wang}, {Wuyts}, \&
  {Xu}}]{santini14}
{Santini}, P., {Maiolino}, R., {Magnelli}, B., {et~al.} 2014, \aap, 562, A30,
  \dodoi{10.1051/0004-6361/201322835}

\bibitem[{{Sargent} {et~al.}(2014){Sargent}, {Daddi}, {B{\'e}thermin},
  {Aussel}, {Magdis}, {Hwang}, {Juneau}, {Elbaz}, \& {da Cunha}}]{sargent14}
{Sargent}, M.~T., {Daddi}, E., {B{\'e}thermin}, M., {et~al.} 2014, \apj, 793,
  19, \dodoi{10.1088/0004-637X/793/1/19}

\bibitem[{{Schinnerer} {et~al.}(2016){Schinnerer}, {Groves}, {Sargent},
  {Karim}, {Oesch}, {Magnelli}, {LeFevre}, {Tasca}, {Civano}, {Cassata}, \&
  {Smol{\v{c}}i{\'c}}}]{schinnerer16}
{Schinnerer}, E., {Groves}, B., {Sargent}, M.~T., {et~al.} 2016, \apj, 833,
  112, \dodoi{10.3847/1538-4357/833/1/112}

\bibitem[{{Schmidt}(1959)}]{schmidt1959}
{Schmidt}, M. 1959, \apj, 129, 243, \dodoi{10.1086/146614}

\bibitem[{{Scoville} {et~al.}(2014){Scoville}, {Aussel}, {Sheth}, {Scott},
  {Sanders}, {Ivison}, {Pope}, {Capak}, {Vand en Bout}, {Manohar},
  {Kartaltepe}, {Robertson}, \& {Lilly}}]{scoville14}
{Scoville}, N., {Aussel}, H., {Sheth}, K., {et~al.} 2014, \apj, 783, 84,
  \dodoi{10.1088/0004-637X/783/2/84}

\bibitem[{{Scoville} {et~al.}(2016){Scoville}, {Sheth}, {Aussel}, {Vanden
  Bout}, {Capak}, {Bongiorno}, {Casey}, {Murchikova}, {Koda},
  {{\'A}lvarez-M{\'a}rquez}, {Lee}, {Laigle}, {McCracken}, {Ilbert}, {Pope},
  {Sanders}, {Chu}, {Toft}, {Ivison}, \& {Manohar}}]{scoville16}
{Scoville}, N., {Sheth}, K., {Aussel}, H., {et~al.} 2016, \apj, 820, 83,
  \dodoi{10.3847/0004-637X/820/2/83}

\bibitem[{{Scoville} {et~al.}(2017){Scoville}, {Lee}, {Vanden Bout},
  {Diaz-Santos}, {Sanders}, {Darvish}, {Bongiorno}, {Casey}, {Murchikova},
  {Koda}, {Capak}, {Vlahakis}, {Ilbert}, {Sheth}, {Morokuma-Matsui}, {Ivison},
  {Aussel}, {Laigle}, {McCracken}, {Armus}, {Pope}, {Toft}, \&
  {Masters}}]{scoville17}
{Scoville}, N., {Lee}, N., {Vanden Bout}, P., {et~al.} 2017, \apj, 837, 150,
  \dodoi{10.3847/1538-4357/aa61a0}

\bibitem[{{Seko} {et~al.}(2016{\natexlab{a}}){Seko}, {Ohta}, {Yabe},
  {Hatsukade}, {Akiyama}, {Iwamuro}, {Tamura}, \& {Dalton}}]{seko16_alma}
{Seko}, A., {Ohta}, K., {Yabe}, K., {et~al.} 2016{\natexlab{a}}, \apj, 819, 82,
  \dodoi{10.3847/0004-637X/819/1/82}

\bibitem[{{Seko} {et~al.}(2016{\natexlab{b}}){Seko}, {Ohta}, {Yabe},
  {Hatsukade}, {Akiyama}, {Tamura}, {Iwamuro}, \& {Dalton}}]{seko16}
---. 2016{\natexlab{b}}, \apj, 833, 53, \dodoi{10.3847/1538-4357/833/1/53}

\bibitem[{{Shapley} {et~al.}(2020){Shapley}, {Cullen}, {Dunlop}, {McLure},
  {Kriek}, {Reddy}, \& {Sand ers}}]{shapley20}
{Shapley}, A.~E., {Cullen}, F., {Dunlop}, J.~S., {et~al.} 2020, arXiv e-prints,
  arXiv:2009.10091.
\newblock \doarXiv{2009.10091}

\bibitem[{{Sobral} {et~al.}(2014){Sobral}, {Best}, {Smail}, {Mobasher},
  {Stott}, \& {Nisbet}}]{sobral14}
{Sobral}, D., {Best}, P.~N., {Smail}, I., {et~al.} 2014, \mnras, 437, 3516,
  \dodoi{10.1093/mnras/stt2159}

\bibitem[{{Sobral} {et~al.}(2013){Sobral}, {Smail}, {Best}, {Geach}, {Matsuda},
  {Stott}, {Cirasuolo}, \& {Kurk}}]{sobral13}
{Sobral}, D., {Smail}, I., {Best}, P.~N., {et~al.} 2013, \mnras, 428, 1128,
  \dodoi{10.1093/mnras/sts096}

\bibitem[{{Speagle} {et~al.}(2014){Speagle}, {Steinhardt}, {Capak}, \&
  {Silverman}}]{speagle14}
{Speagle}, J.~S., {Steinhardt}, C.~L., {Capak}, P.~L., \& {Silverman}, J.~D.
  2014, \apjs, 214, 15, \dodoi{10.1088/0067-0049/214/2/15}

\bibitem[{{Sugahara} {et~al.}(2017){Sugahara}, {Ouchi}, {Lin}, {Martin}, {Ono},
  {Harikane}, {Shibuya}, \& {Yan}}]{sugahara17}
{Sugahara}, Y., {Ouchi}, M., {Lin}, L., {et~al.} 2017, \apj, 850, 51,
  \dodoi{10.3847/1538-4357/aa956d}

\bibitem[{{Suzuki} {et~al.}(2017){Suzuki}, {Kodama}, {Onodera}, {Shimakawa},
  {Hayashi}, {Tadaki}, {Koyama}, {Tanaka}, {Sobral}, {Smail}, {Best},
  {Khostovan}, {Minowa}, \& {Yamamoto}}]{suzuki17}
{Suzuki}, T.~L., {Kodama}, T., {Onodera}, M., {et~al.} 2017, \apj, 849, 39,
  \dodoi{10.3847/1538-4357/aa8df3}

\bibitem[{{Tacchella} {et~al.}(2020){Tacchella}, {Forbes}, \&
  {Caplar}}]{tacchella20}
{Tacchella}, S., {Forbes}, J.~C., \& {Caplar}, N. 2020, arXiv e-prints,
  arXiv:2006.09382.
\newblock \doarXiv{2006.09382}

\bibitem[{{Tacconi} {et~al.}(2010){Tacconi}, {Genzel}, {Neri}, {Cox}, {Cooper},
  {Shapiro}, {Bolatto}, {Bouch{\'e}}, {Bournaud}, {Burkert}, {Combes},
  {Comerford}, {Davis}, {Schreiber}, {Garcia-Burillo}, {Gracia-Carpio}, {Lutz},
  {Naab}, {Omont}, {Shapley}, {Sternberg}, \& {Weiner}}]{tacconi10}
{Tacconi}, L.~J., {Genzel}, R., {Neri}, R., {et~al.} 2010, \nat, 463, 781,
  \dodoi{10.1038/nature08773}

\bibitem[{{Tacconi} {et~al.}(2013){Tacconi}, {Neri}, {Genzel}, {Combes},
  {Bolatto}, {Cooper}, {Wuyts}, {Bournaud}, {Burkert}, {Comerford}, {Cox},
  {Davis}, {F{\"o}rster Schreiber}, {Garc{\'{\i}}a-Burillo}, {Gracia-Carpio},
  {Lutz}, {Naab}, {Newman}, {Omont}, {Saintonge}, {Shapiro Griffin}, {Shapley},
  {Sternberg}, \& {Weiner}}]{tacconi13}
{Tacconi}, L.~J., {Neri}, R., {Genzel}, R., {et~al.} 2013, \apj, 768, 74,
  \dodoi{10.1088/0004-637X/768/1/74}

\bibitem[{{Tacconi} {et~al.}(2018){Tacconi}, {Genzel}, {Saintonge}, {Combes},
  {Garc{\'{\i}}a-Burillo}, {Neri}, {Bolatto}, {Contini}, {F{\"o}rster
  Schreiber}, {Lilly}, {Lutz}, {Wuyts}, {Accurso}, {Boissier}, {Boone},
  {Bouch{\'e}}, {Bournaud}, {Burkert}, {Carollo}, {Cooper}, {Cox}, {Feruglio},
  {Freundlich}, {Herrera-Camus}, {Juneau}, {Lippa}, {Naab}, {Renzini},
  {Salome}, {Sternberg}, {Tadaki}, {{\"U}bler}, {Walter}, {Weiner}, \&
  {Weiss}}]{tacconi18}
{Tacconi}, L.~J., {Genzel}, R., {Saintonge}, A., {et~al.} 2018, \apj, 853, 179,
  \dodoi{10.3847/1538-4357/aaa4b4}

\bibitem[{{Tan} {et~al.}(2014){Tan}, {Daddi}, {Magdis}, {Pannella}, {Sargent},
  {Riechers}, {B{\'e}thermin}, {Bournaud}, {Carilli}, {da Cunha},
  {Dannerbauer}, {Dickinson}, {Elbaz}, {Gao}, {Hodge}, {Owen}, \&
  {Walter}}]{tan14}
{Tan}, Q., {Daddi}, E., {Magdis}, G., {et~al.} 2014, \aap, 569, A98,
  \dodoi{10.1051/0004-6361/201423905}

\bibitem[{{Taylor}(2005)}]{topcat}
{Taylor}, M.~B. 2005, in Astronomical Society of the Pacific Conference Series,
  Vol. 347, Astronomical Data Analysis Software and Systems XIV, ed.
  P.~{Shopbell}, M.~{Britton}, \& R.~{Ebert}, 29

\bibitem[{{Tomczak} {et~al.}(2016){Tomczak}, {Quadri}, {Tran}, {Labb{\'e}},
  {Straatman}, {Papovich}, {Glazebrook}, {Allen}, {Brammer}, {Cowley},
  {Dickinson}, {Elbaz}, {Inami}, {Kacprzak}, {Morrison}, {Nanayakkara},
  {Persson}, {Rees}, {Salmon}, {Schreiber}, {Spitler}, \&
  {Whitaker}}]{tomczak16}
{Tomczak}, A.~R., {Quadri}, R.~F., {Tran}, K.-V.~H., {et~al.} 2016, \apj, 817,
  118, \dodoi{10.3847/0004-637X/817/2/118}

\bibitem[{{Torrey} {et~al.}(2019){Torrey}, {Vogelsberger}, {Marinacci},
  {Pakmor}, {Springel}, {Nelson}, {Naiman}, {Pillepich}, {Genel}, {Weinberger},
  \& {Hernquist}}]{torrey19}
{Torrey}, P., {Vogelsberger}, M., {Marinacci}, F., {et~al.} 2019, \mnras, 484,
  5587, \dodoi{10.1093/mnras/stz243}

\bibitem[{{Troncoso} {et~al.}(2014){Troncoso}, {Maiolino}, {Sommariva},
  {Cresci}, {Mannucci}, {Marconi}, {Meneghetti}, {Grazian}, {Cimatti},
  {Fontana}, {Nagao}, \& {Pentericci}}]{troncoso14}
{Troncoso}, P., {Maiolino}, R., {Sommariva}, V., {et~al.} 2014, \aap, 563, A58,
  \dodoi{10.1051/0004-6361/201322099}

\bibitem[{{Walter} {et~al.}(2016){Walter}, {Decarli}, {Aravena}, {Carilli},
  {Bouwens}, {da Cunha}, {Daddi}, {Ivison}, {Riechers}, {Smail}, {Swinbank},
  {Weiss}, {Anguita}, {Assef}, {Bacon}, {Bauer}, {Bell}, {Bertoldi}, {Chapman},
  {Colina}, {Cortes}, {Cox}, {Dickinson}, {Elbaz}, {G{\'o}nzalez-L{\'o}pez},
  {Ibar}, {Inami}, {Infante}, {Hodge}, {Karim}, {Le Fevre}, {Magnelli}, {Neri},
  {Oesch}, {Ota}, {Popping}, {Rix}, {Sargent}, {Sheth}, {van der Wel}, {van der
  Werf}, \& {Wagg}}]{walter16}
{Walter}, F., {Decarli}, R., {Aravena}, M., {et~al.} 2016, \apj, 833, 67,
  \dodoi{10.3847/1538-4357/833/1/67}

\bibitem[{{Whitaker} {et~al.}(2012){Whitaker}, {van Dokkum}, {Brammer}, \&
  {Franx}}]{whitaker12}
{Whitaker}, K.~E., {van Dokkum}, P.~G., {Brammer}, G., \& {Franx}, M. 2012,
  \apjl, 754, L29, \dodoi{10.1088/2041-8205/754/2/L29}

\bibitem[{{Wiklind} {et~al.}(2019){Wiklind}, {Ferguson}, {Guo}, {Koo},
  {Kocevski}, {Mobasher}, {Brammer}, {Kassin}, {Koekemoer}, {Giavalisco},
  {Papovich}, {Ravindranath}, {Faber}, {Freundlich}, \& {de Mello}}]{wiklind19}
{Wiklind}, T., {Ferguson}, H.~C., {Guo}, Y., {et~al.} 2019, \apj, 878, 83,
  \dodoi{10.3847/1538-4357/ab1089}

\bibitem[{{Yabe} {et~al.}(2015){Yabe}, {Ohta}, {Akiyama}, {Iwamuro}, {Tamura},
  {Yuma}, {Dalton}, \& {Lewis}}]{yabe15_apj}
{Yabe}, K., {Ohta}, K., {Akiyama}, M., {et~al.} 2015, \apj, 798, 45,
  \dodoi{10.1088/0004-637X/798/1/45}

\bibitem[{{Zahid} {et~al.}(2014){Zahid}, {Dima}, {Kudritzki}, {Kewley},
  {Geller}, {Hwang}, {Silverman}, \& {Kashino}}]{zahid_apj791}
{Zahid}, H.~J., {Dima}, G.~I., {Kudritzki}, R.-P., {et~al.} 2014, \apj, 791,
  130, \dodoi{10.1088/0004-637X/791/2/130}

\end{thebibliography}
%\bibliographystyle{aasjournal}



\end{document}